\documentclass[letterpaper,british,english]{emulateapj}
\usepackage{ae,aecompl}
\usepackage[T1]{fontenc}
\usepackage[latin9]{inputenc}
\usepackage{array}
\usepackage{float}
\usepackage{units}
\usepackage{multirow}
\usepackage{amsbsy}
\usepackage{amstext}
\usepackage{amssymb}
\usepackage{graphicx}

\makeatletter

\pdfpageheight\paperheight
\pdfpagewidth\paperwidth

\providecommand{\tabularnewline}{\\}

\slugcomment{}
\shorttitle{}
\shortauthors{}

\usepackage{amsmath}

\usepackage{enumitem}  
\setlist{nolistsep}

\usepackage{graphicx}

\makeatother

\usepackage{babel}
\begin{document}

\title{How to Distinguish between Cloudy Mini-Neptunes and\\
Water/Volatile-Dominated Super-Earths}

\author{Björn Benneke and Sara Seager}

\affil{Department of Earth, Atmospheric and Planetary Sciences, Massachusetts
Institute of Technology, Cambridge, MA 02139; USA}

\email{bbenneke@mit.edu}
\begin{abstract}
One of the most profound questions about the newly discovered class
of low-density super-Earths is whether these exoplanets are predominately
$\mathrm{H_{2}}$-dominated mini-Neptunes or volatile-rich worlds
with gas envelopes dominated by $\mathrm{H_{2}O}$, $\mathrm{CO_{2}}$,
$\mathrm{CO}$, $\mathrm{CH_{4}}$, or $\mathrm{N_{2}}$. Transit
observations of the super-Earth GJ~1214b rule out cloud-free $\mathrm{H_{2}}$-dominated
scenarios, but are not able to determine whether the lack of deep
spectral features is due to high-altitude clouds or the presence of
a high mean molecular mass atmosphere. 

Here, we demonstrate that one can unambiguously distinguish between
cloudy mini-Neptunes and volatile-dominated worlds based on the differences
in the wing steepness and relative depths of water absorption features
in moderate-resolution NIR transmission spectra ($R\sim100$). In
a numerical retrieval study, we show for GJ~1214b that an unambiguous
distinction between a cloudy $\mathrm{H_{2}}$-dominated atmosphere
and cloud-free $\mathrm{H_{2}O}$ atmosphere will be possible if the
uncertainties in the spectral transit depth measurements can be reduced
by a factor of $\sim3$ compared to the published \textit{HST~WFC3}
and \textit{VLT} transit observations by \citet{berta_flat_2012}
and \citet{bean_ground-based_2010}. We argue that the required precision
for the distinction may be achievable with currently available instrumentation
by stacking $10-15$ repeated transit observations. We provide a scaling
law that scales our quantitative results to other transiting super-Earths
and Neptunes such as HD~97658b, 55~Cnc~e, and GJ~436b. 

The analysis in this work is performed using an improved version of
our Bayesian atmospheric retrieval framework. The new framework not
only constrains the gas composition and cloud/haze parameters, but
also determines our confidence in having detected molecules and cloud/haze
species through Bayesian model comparison. Using the Bayesian tool,
we demonstrate quantitatively that the subtle transit depth variation
in the \citet{berta_flat_2012} data is not sufficient to claim the
detection of water absorption. 
\end{abstract}

\section{Introduction\label{sec:Introduction}}

Super-Earth exoplanets, with masses between 1 and 10 Earth masses,
lie in the intermediate mass range between terrestrial planets and
gas and ice giants in the solar system. Compelling questions arise
as to the composition and nature of these objects and whether they
are capable of harboring life. According to theoretical studies,
\citep[e.g.,][]{seager_massradius_2007,rogers_framework_2010,nettelmann_thermal_2011}
many super-Earth exoplanets show a bulk density that is high enough
to require a larger ice or rock fraction than the solar system ice
giants, but far too low to be explained by an entirely Earth-like
rocky composition. \citet{rogers_three_2010} showed that their bulk
density may, instead, be explained by either planets that have accreted
and maintained a thick $\mathrm{H_{2}}$/He envelope atop an ice and
rock core or, alternatively, by a new class of ``water worlds''
which contain a large fraction of water or ices in their interior
and are surrounded by a dense water vapor atmosphere (\citet{kuchner_volatile-rich_2003}
and \citet{leger_new_2004}).

One way of answering questions about the nature and habitability of
super-Earths is to identify their atmospheric thicknesses and molecular
compositions by observing their transmission and/or thermal emission
spectra. \citet{miller-ricci_atmospheric_2009} showed that cloud-free,
hydrogen-dominated atmospheres would display absorption features in
the transmission spectrum that are several times larger than those
expected for atmospheres dominated by water vapor, $\mathrm{CO_{2}}$,
$\mathrm{CO}$, $\mathrm{CH_{4}}$, or $\mathrm{N_{2}}$ due to the
lower mean molecular mass and resulting larger scale height.

Many observational attempts to detect an atmosphere around the super-Earth
GJ~1214b and characterize its composition have been made, but the
individual observational data sets were found to be insufficient to
identify the presence of atmospheric features to within observational
uncertainty \citep{bean_ground-based_2010,bean_optical_2011,desert_observational_2011,carter_transit_2011,crossfield_high-resolution_2011,berta_gj1214_2011,berta_flat_2012,de_mooij_optical_2012,teske_optical_2013}.
An initial finding of a difference in the transit depths between
the J and Ks band by \citet{croll_broadband_2011} could not be confirmed
by \citet{bean_optical_2011}.

The absence of deep features in the transmission spectrum of GJ~1214b
rules out the presence of a cloud-free hydrogen-dominated atmosphere,
but the obtained observational data were shown to be compatible with
high mean molecular mass atmospheres, such as water vapor-dominated
atmospheres, as well as with a hydrogen-dominated atmosphere in the
presence of high altitude clouds \citep{berta_flat_2012}. The interpretation
of the observational data for GJ~1214b revealed the limitations of
the absorption feature depths as a measure of the atmosphere's hydrogen
content in the presence of clouds. Theoretical studies of the atmosphere
loss were conducted attempting to understand the stability of different
atmospheres on highly irradiated super-Earths \citep[e.g.,][]{heng_stability_2012,fortney_framework_2013,kurokawa_atmospheric_2013}.
It appears inevitable, however, that eventually we need observational
proof to understand or confirm the nature of low-density super-Earths.

For general atmospheres that may contain clouds, \citet{benneke_atmospheric_2012}
showed that the unambiguous effect of the mean molecular mass $\mu_{\mathrm{ave}}$
on the transmission spectrum is not the overall depths of the molecular
features, but the gradient \textbf{$\mathrm{d}R_{P,\lambda}/\mathrm{d}\text{\ensuremath{\left(\ln\sigma_{\lambda}\right)}}$}
with which the observed planetary radius $R_{P,\lambda}$ changes
as the atmospheric opacity $\sigma_{\lambda}$ changes across the
spectrum:

\begin{equation}
\mu_{\mathrm{ave}}=\frac{k_{B}T}{g_{\mathrm{p}}}\left(\frac{\mathrm{d}R_{P,\lambda}}{\mathrm{d}\text{\ensuremath{\left(\ln\sigma_{\lambda}\right)}}}\right)^{-1}\times\left(1\pm\frac{\delta T}{T}\right).\label{eq:1}
\end{equation}

The surface gravity $g_{\mathrm{p}}$ can be determined directly from
the transit light curve and radial velocity measurements \citep{winn_transits_2011},
$k_{B}$ is Boltzmann's constant, and the atmospheric temperature
$T$ can be approximated by the equilibrium temperature or modeled
by a radiative-convective model. The term $\left(1\pm\delta T/T\right)$
accounts for the inherent uncertainty on the mean molecular mass due
to the uncertainty, $\delta T$, in estimating the temperature $T$
at the planetary radius $R_{P,\lambda}$. 

In practice, an estimate of the mean molecular mass can be determined
at visible or NIR wavelengths by measuring the steepness of molecular
feature wings or comparing the relative transit depths in two or more
absorption features of the same molecule. The decrease in opacity
$\sigma_{\lambda}$ from the center of the absorption features to
wing can be modeled based on molecular line list of the absorber.

Measuring the slope of gaseous Rayleigh scattering signature at UV/visible
wavelengths also provide constraints on the mean molecular mass; however,
cloud and haze opacities can mask the signature of gaseous Rayleigh
scattering and the distinction between a shallow slope due to a high
mean molecular mass or clouds may be difficult.

Equation (\ref{eq:1}) demonstrates that, if sufficient observations
of the transmission spectrum are available, the mean molecular mass
can determined to the same relative precision as that at which we
are able to estimate the atmospheric temperature $T$. Since the mean
molecular mass varies by a factor of $\sim8$ or more between hydrogen-rich
atmospheres and atmospheres dominated by water vapor or other volatiles,
we can distinguish between cloudy hydrogen-rich and water-rich atmospheres
even if the temperature $T$ is known only with an uncertainty of
several tens of percent.

The ability to measure the mean molecular mass at NIR wavelengths
is of particular interest for the near-term characterizations of super-Earths
because planets orbiting small M-dwarfs are the ones with the strongest
transit signatures and are most amenable to study at infrared wavelengths
due to the low stellar flux of M-dwarfs at short wavelengths. In
this work, we demonstrate using quantitative simulations that NIR
transit observations near the peak of the stellar spectrum of M-stars
provide a practical means to distinguish between cloudy hydrogen-dominated
atmospheres and atmospheres dominated by water vapor or other volatiles.
 We determine what precision in the transit depth measurements is
required (1) to detect the absorption features in the transmission
spectrum of a water-dominated super-Earth and (2) to unambiguously
distinguish between water-dominated atmospheres and hydrogen-dominated
atmospheres based on the different wing steepnesses of the water absorption
bands.

The paper outline is as follows. In Section~2, we describe the Bayesian
framework used to identify the presence of molecular species, quantify
the statistical significances of molecular detections, and constrain
the abundance of the molecular species in the atmosphere. Section~2
presents a quantitative picture of what observations and noise levels
are required for the super-Earth GJ~1214b to detect the presence
of water vapor with high confidence as well as to distinguish between
cloudy, hydrogen-dominated atmospheres and water-dominated atmospheres.
In Section 4, we present a retrieval analysis of the spectral observations
by \textit{HST WFC3} spectrum of GJ~1214b by \citet{berta_flat_2012}.
Section~5 provides a scaling law to scale the quantitative results
to super-Earth and Neptunes such as HD~97658b, 55~Cnc~e, and GJ~436b.
We also discuss the near-term feasibility of the proposed study. Finally,
we present our summary and conclusions in Section~6.

\section{Methods\label{sec:Methods}}

The main goal of this work is to determine the level of precision
in NIR super-Earth transmission spectra that is required to unambiguously
distinguish between atmospheres dominated by hydrogen/helium and atmospheres
dominated by water vapor or other volatiles such as $\mathrm{CO_{2}}$,
$\mathrm{CO}$, $\mathrm{CH_{4}}$, or $\mathrm{N_{2}}$. We address
this question by computing \textit{synthetic} transit observations
derived from super-Earth model transmission spectra and analyzing
them using a Bayesian atmospheric retrieval framework.   

The Bayesian retrieval method employed in this work builds upon ideas
introduced in \citet{benneke_atmospheric_2012}, but was extended
by a Bayesian model comparison framework that enables one to rationally
decide which molecular species and types of aerosols are present in
the atmosphere. We employ the nested sampling technique \citep{skilling_nested_2004,feroz_multimodal_2008}
to efficiently compute and compare the Bayesian evidences of retrieval
models with different complexities. Once a retrieval model is identified
that is adequate in light of the data, the constraints on atmospheric
parameters are inferred from the joint posterior probability distribution
obtained as a by-product of the nested sampling calculation.

\subsection{Atmospheric ``Forward'' Model\label{sub:Atmosphere-Forward-Models}}

We use the 1D exoplanetary atmosphere ``forward'' model originally
described in \citet{benneke_atmospheric_2012} to compute model transmission
spectra and synthetic observations. Our model uses line-by-line radiative
transfer in local thermodynamic equilibrium, hydrostatic equilibrium,
and a temperature-pressure profile consistent with the atmospheric
composition.

The input to the atmospheric ``forward'' model is an adaptable set
of free model parameters describing the gas composition and aerosol
properties in the model atmosphere. A difference between this model
and the model used in \citet{benneke_atmospheric_2012} is that the
aerosols considered in this work encompass small particle hazes as
well as larger cloud particles. Absorption and scattering of the particles
are modeled using Mie theory for spherical particles \citep{hansen_light_1974}.
The complex refractive indices of the condensate species Potassium
Chloride (KCl) or Zinc Sulfide (ZnS) used in this work for GJ~1214b
are taken from \citet{querry_optical_1987}.

Synthetic observations are generated by first specifying the chemical
composition and aerosol opacities of the individual atmospheric layers.
 Synthetic observations are then derived from the model transmission
spectra by integrating the model spectra over flat instrument response
functions with spectral coverage equivalent to the spectral points
from the \textit{VLT }observations by \citet{bean_ground-based_2010}
and the \textit{HST WFC3} observations by \citet{berta_flat_2012}.
Gaussian noise ranging from $\sim180\,\mathrm{ppm}$ down to $35\,\mathrm{ppm}$
is added to the data to generate synthetic observations of different
precisions. 

The two main scenarios we aim to unambiguously distinguish in this
work are water- or volatile-dominated atmospheres and hydrogen-rich
atmospheres with high-altitude clouds. In addition, we would like
to contrast these two scenarios from a flat transmission spectra which
would be observed if the planet lacks a gaseous atmosphere or clouds
are present at extremely high altitude such that all molecular features
are muted. As representative example for these three scenarios, we
select a cloud-free water-dominated atmosphere composed (95\% $\mathrm{H_{2}O}$
+ 5\% $\mathrm{CO_{2}}$), a methane-depleted solar metallicity atmosphere
with a cloud deck at 10 mbar, and a featureless spectrum. The level
of the high altitude clouds in the solar composition scenario is chosen
such that the overall depths of the absorption features resemble that
of a cloud-free water-rich atmosphere. Methane is removed from the
atmospheric scenario to demonstrate that the distinction between water/volatile-dominated
scenarios and hydrogen-dominated scenarios is possible based solely
on the wing steepnesses and relative sizes of the water features.

The atmospheric scenarios for the synthetic observations are chosen
to be plausible scenarios, but are not calculated from a fully self-consistent
model. The goal of this work is to demonstrate the retrieval method
for exoplanetary atmospheres for which we do not have a full understanding
of the physical and chemical behavior prior to the observations.

\subsection{Bayesian Atmospheric Retrieval}

Atmospheric retrieval aims at solving the inverse problem of atmospheric
``forward'' modeling: ``Given an observed planetary spectrum, what
are the properties of the planetary atmosphere?'' The atmospheric
retrieval problem can be solved using \textit{Bayesian parameter estimation}
by first choosing a retrieval model that defines the hypothesis space
in the form of a set of atmospheric parameters and then deriving constraints
on those parameters \citep{benneke_atmospheric_2012}. 

In choosing the atmospheric retrieval model, however, questions arise
as to how much complexity and how many free parameters should be included
in the retrieval model. In this work, we introduce \textit{Bayesian
model comparison }to determine which molecular gases and types of
aerosols can be inferred from the data and need to be included in
the analysis. The approach enables us to rationally adapt the complexity
of the retrieval model to the available data. Exquisite observational
data with high signal-to-noise ratio (S/N) and high spectral resolution,
as are currently only available for solar system planets, allow the
inference of the detailed abundance profiles of the molecular species,
the temperature structure, and the presence of particles in the atmospheres.
At lower S/N and sparse spectral coverage, however, complex models
with large numbers of free parameters would overfit the available
data and introduce strong degeneracies. It is therefore necessary
to adjust the number of free parameters in the retrieval model according
to the amount and precision of observational data available.

\subsubsection{Bayesian Estimation of Atmospheric Parameters}

An atmospheric retrieval model $M_{i}$ in our Bayesian framework
is defined as a set of atmospheric parameters $\boldsymbol{\theta}=[\theta_{1},\,\ldots,\,\theta_{N}]$
and their joint prior probability distribution $\pi\text{\ensuremath{\left(\boldsymbol{\theta}|M_{i}\right)}}$.
The atmospheric parameters considered in the retrieval models for
transmission spectra are the mole fractions (or volume mixing ratios)
of the atmospheric gases, the haze or cloud top pressure, and the
radius at a reference pressure level. Since we generally have little
prior knowledge of the state of the exoplanetary atmosphere, we assign
non-informative (uniform) priors to the atmospheric parameters (Table
\ref{tab:Prior-probability-}). We include the Planetary Bond albedo
as a free parameters to account for the uncertainty in the temperature
structure introduced by our ignorance about how much of solar radiation
is reflected by the planet.

\begin{table*}[t]
\begin{centering}
{\small }%
\begin{tabular}{>{\raggedright}p{4.5cm}l>{\raggedright}p{3cm}c>{\centering}p{3cm}}
\hline 
{\small Parameter} &  & {\small Prior} & {\small Lower Bound} & {\small Upper Bound}\tabularnewline
\hline 
\hline 
{\small Planet-to-star radius ratio} & {\small $R_{\mathrm{P}}/R_{*}$} & {\small Uniform on log-scale} & {\small $R_{\mathrm{Mars}}/R_{*}$} & {\small $1$}\tabularnewline
\multirow{1}{4.5cm}{{\small Centered-log-ratio transform\linebreak  of mole fractions}} & \multirow{1}{*}{{\small $\boldsymbol{\boldsymbol{\xi}}$}} & \multirow{1}{3cm}{{\small Uniform in $\boldsymbol{\xi}$-space}} & \multicolumn{2}{>{\centering}p{6cm}}{{\small bound by condition that the mole fractions of all present
molecular species are greater than $1\,\mathrm{ppb}$ $^{1}$}}\tabularnewline
{\small Cloud-top pressure} & {\small $P_{\mathrm{surf}}$} & {\small Uniform on log-scale} & {\small $1\,\mathrm{\mu bar}$} & {\small 100~bar}\tabularnewline
{\small Planetary Bond albedo} & {\small $A_{\mathrm{B}}$} & {\small Uniform} & {\small 0} & {\small 1}\tabularnewline
\hline 
\end{tabular}
\par\end{centering}{\small \par}

\caption{Prior probability of atmospheric retrieval parameters. $^{1}$The
effect of molecular species with mole fractions below 1~ppb is negligible
given currently available observations. \label{tab:Prior-probability-} }
\end{table*}

Once a retrieval model is identified that is adequate in light of
the data, we can infer the constraints on the model parameters by
computing their joint posterior probability distribution $p\text{\ensuremath{\left(\boldsymbol{\theta}|M_{i},\boldsymbol{D}\right)}}$
\citep{benneke_atmospheric_2012}. Using Bayes' Law, we write the
posterior distribution of the atmospheric parameters as

\begin{equation}
p\text{\ensuremath{\left(\boldsymbol{\theta}|M_{i},\boldsymbol{D}\right)}}=\frac{\pi\text{\ensuremath{\left(\boldsymbol{\boldsymbol{\theta}}|M_{i}\right)}}\mathcal{L}\ensuremath{\left(\boldsymbol{D}|M_{i},\boldsymbol{\theta}\right)}}{\mathcal{Z}\left(\boldsymbol{D}|M_{i}\right)},\label{eq:Bayesian-1-2-1}
\end{equation}

where $\mathcal{L}\ensuremath{\left(\boldsymbol{D}|M_{i},\boldsymbol{\theta}\right)}$
is the likelihood function and $\mathcal{Z}\left(\boldsymbol{D}|M_{i}\right)$
is the Bayesian evidence of the model $M_{i}$. The likelihood function
$\mathcal{L}\ensuremath{\left(\boldsymbol{D}|M_{i},\boldsymbol{\theta}\right)}$
is the probability of observing the data $\boldsymbol{D}$, given
that the atmospheric parameters are $\boldsymbol{\theta}$. The likelihood
function is modeled using the atmospheric \textquotedbl{}forward\textquotedbl{}
model (Section~\ref{sub:Atmosphere-Forward-Models}). For independent
Gaussian errors in the spectral observations, the likelihood function
is

\begin{equation}
\mathcal{L}\ensuremath{\left(\boldsymbol{D}|M_{i},\boldsymbol{\theta}\right)}=\prod_{k=1}^{N}\frac{1}{\sigma_{k}\sqrt{2\pi}}\exp\left\{ -\frac{\left[D_{k,obs}-D_{k,model}\left(\boldsymbol{\theta}\right)\right]^{2}}{2\sigma_{k}^{2}}\right\} ,\label{eq:likelihoodfct}
\end{equation}

where $D_{k,obs}$ is the $k$-th observational data point in the
spectrum, $D_{k,model}\left(\boldsymbol{\theta}\right)$ is the model
prediction for the $k$-th data point given a set of atmospheric parameters
$\boldsymbol{\theta}$, and $N$ is the total number of data points
in the observed spectrum. The denominator $ $$\mathcal{Z}\left(\boldsymbol{D}|M_{i}\right)$
is the Bayesian evidence and merely a normalization constant in Equation
(\ref{eq:Bayesian-1-2-1}). It is, however, central to Bayesian model
comparison as described next.

\subsubsection{Identification of Gases and Aerosols using Bayesian Model Comparison\label{sub:Bayesian-Retrieval-Model}}

One of the questions that arises when interpreting spectra of planetary
atmospheres is which molecular species and aerosol types can be inferred
from a given data set. Spectral signatures of molecular gases or
aerosols are often hidden in the noisy observations of exoplanets.
In addition, strong overlap between absorption features of different
molecules and the lack of a flat continuum for thick atmospheres may
further complicate the identification of individual gases.

\textit{Bayesian model comparison} offers a rational way of determining
which atmospheric species can be inferred from the data. It has been
widely used in cosmology where noisy data and little prior knowledge
require sophisticated statistical tools \citep[see][for excellent reviews]{trotta_bayes_2008,hobson_bayesian_2009}.
In the exoplanet context, Bayesian model comparison was employed by
\citet{gregory_bayesian_2007} to detect planets in radial velocity
data.

In our atmospheric retrieval application, we quantify our confidence
in having detected the presence of a particular atmospheric constituent
$m$ by computing the posterior odds ratio (or Bayes factor) between
a simpler retrieval model that neglect the presence of the species
$m$ and a more complex retrieval model that includes an additional
parameter to describe the abundances of the constituent $m$. If the
posterior odds ratio is strongly in favor of the more complex model,
we can safely conclude that the atmospheric constituent is present. 

In practice, the required comparisons for all relevant atmospheric
constituents are achieved by computing the Bayesian evidence for one
retrieval model that covers the full prior hypothesis space and comparing
it to a list of retrieval models for which individual molecular species
and types of aerosols were removed from the prior hypothesis space.
The approach ensures that a high confidence in the presence of a particular
atmospheric constituent is reported only if no other constituent in
the prior hypothesis space can have resulted in the observed data.

\paragraph{Bayesian evidence}

The quantity central to all Bayesian model comparisons is the Bayesian
evidence defined as

\begin{equation}
\mathcal{Z}\left(\boldsymbol{D}|M_{i}\right)=\int_{all\,\boldsymbol{\theta}}\,\pi(\boldsymbol{\theta}|M_{i})\,\mathcal{L}\ensuremath{\left(\boldsymbol{D}|M_{i},\boldsymbol{\theta}\right)}\,\mathrm{d}^{N}\boldsymbol{\theta}.\label{eq:BayesianEvidence}
\end{equation}

The Bayesian evidence quantifies the adequacy of a retrieval model
$M_{i}$, specified by a set of $N$ atmospheric parameters $\boldsymbol{\theta}=[\theta_{1},\,\ldots,\,\theta_{N}]$
and their prior probability distribution $\pi(\boldsymbol{\theta}|M_{i})$,
in the light of the observational data $\boldsymbol{D}$. Comparing
the Bayesian evidences of alternative retrieval models with parameterizations
of lower and higher complexities enables one to assess which atmospheric
constituents and effects are important and which model complexities
ought to be removed. 

The underlying idea of the Bayesian model comparison is thereby that
additional complexity of the parameter space ought to be avoided whenever
a simpler model provides an adequate representation of the data. While
it is obvious that a best fit of a retrieval model with more free
parameters will always be better than (or at least as good as) the
best fit of a model with fewer parameters, the Bayesian evidence only
favors the more complex model if the improvement in the best fit due
to additional parameters is large enough to overcome the so-called
Occam penalty for the more complex model parameter space.

\paragraph{Posterior odds ratio / Bayes factor}

Using Bayes' law we can express the posterior odds ratio between two
alternative models $\mathrm{M_{\mathrm{``m\, present"}}}$ and as
$\mathrm{M_{\mathrm{``m\, not\, present"}}}$

\begin{equation}
\frac{p\left(\mathrm{M_{\mathrm{``\mathit{m}\, is\, present"}}}|\boldsymbol{D}\right)}{p\left(\mathrm{M_{``\mathit{m}\, not\, present"}}|\boldsymbol{D}\right)}=B_{m}\frac{p\left(\mathrm{M_{\mathrm{``\mathit{m}\, is\, present"}}}\right)}{p\left(M_{\mathrm{``\mathit{m}\, not\, present"}}\right)}
\end{equation}

where the Bayes factor $B_{m}$ is the ratio of the model's Bayesian
evidences

\begin{equation}
B_{m}=\frac{\mathcal{Z}\left(\boldsymbol{D}|M_{``m\,\mathrm{present"}}\right)}{\mathcal{Z}\left(\boldsymbol{D}|M_{``\mathit{m}\, not\, present"}\right)},\label{eq:BayesFactor}
\end{equation}

and $p\left(\mathrm{M_{\mathrm{``m\, present"}}}\right)/p\left(M_{\mathrm{``m\, not\, present"}}\right)$
is the prior odds ratio, which we assume to be 1 in the absence of
prior information.

The posterior probability for the present species $m$ is

\begin{equation}
p\left(\mathrm{``\mathit{m}\, is\, present"}|\boldsymbol{D}\right)=\frac{B_{m}}{1+B_{m}}.
\end{equation}
The Bayes factor $B_{m}$ plays a pivotal role in a Bayesian model
comparison. A value of $B_{m}>1$ means that the data provide support
in favor of the presence of the atmospheric constituent $m$. Bayes
factors are generally interpreted against the Jeffrey scale (ref).
Values of the odds of 3:1, 12:1, and 150:1 represent weak, moderate,
and strong support in favor of the presence of additional molecules
(Table~\ref{tab:Translation-table-between}).

\paragraph{Calibration between Bayesian and Frequentist Detections}

The increase in availability of computational resources over the last
decades has led to a widespread of Bayesian techniques in the analysis
of astrophysical observations. Parameter estimation problems are frequently
solved in a Bayesian framework using Markov Chain Monte Carlo (MCMC)
techniques. Yet, the significance of detections of atmospheric absorption
features is generally still reported in terms of the frequentist measures
of confidence such as the p-value or the ``sigma''-significance.
As a result, many scientists are accustomed to the interpretation
of sigma significance, but not Bayes factors.

A useful calibration between the Bayes factor and the frequentist
measures of confidence is provided by the expression 

\begin{equation}
B_{m}\leq-\frac{1}{\mathrm{e}\cdot\rho\cdot\ln\rho},\label{eq:translation}
\end{equation}

where e is the exponential of one and $\rho$ is the p-value\citep{sellke_calibration_2001}.
The p-value, in turn, can be related to the sigma significance $n_{\sigma}$
using the expression

\begin{equation}
\rho=1-\mathrm{erf}\left(\frac{n_{\sigma}}{\sqrt{2}}\right),
\end{equation}

where erf is the error function. Representative values for a range
of confidence values are listed in Table~\ref{tab:Translation-table-between}.
Equation (\ref{eq:translation}) presents an upper bound on the Bayes
factor, e.g., a Bayes factor $B_{m}=21$ corresponds, at least, to
a $3.0\sigma$ detection. Equation (\ref{eq:translation}) is valid
for $\rho<\mathrm{e}^{-1}$ and under the assumption of a mild principle
of indifference as to the value of the added parameter in the more
complex model \citep{sellke_calibration_2001}.

\begin{table}[t]
\begin{centering}
{\small }%
\begin{tabular}{ccccc}
\hline 
{\small $p$-value} & {\small $B_{m}$} & {\small $\ln B_{m}$} & {\small ``sigma''} & {\small interpretation}\tabularnewline
\hline 
\hline 
{\small 0.05} & {\small 2.5} & {\small 0.9} & {\small 2.0$\sigma$} & \tabularnewline
{\small 0.04} & {\small 2.9} & {\small 1.0} & {\small 2.1$\sigma$} & {\small ``weak'' detection}\tabularnewline
{\small 0.01} & {\small 8.0} & {\small 2.1} & {\small 2.6$\sigma$} & \tabularnewline
{\small 0.006} & {\small 12} & {\small 2.5} & {\small 2.7$\sigma$} & {\small ``moderate'' detection}\tabularnewline
{\small 0.003} & {\small 21} & {\small 3.0} & {\small 3.0$\sigma$} & \tabularnewline
{\small 0.001} & {\small 53} & {\small 4.0} & {\small 3.3$\sigma$} & \tabularnewline
{\small 0.0003} & {\small 150} & {\small 5.0} & {\small 3.6$\sigma$} & {\small ``strong'' detection}\tabularnewline
{\small $6\cdot10^{-7}$} & {\small 43000} & {\small 11} & {\small 5.0$\sigma$} & \tabularnewline
\hline 
\end{tabular}
\par\end{centering}{\small \par}

\caption{Translation table between frequentist significance values ($p$-values)
and the Bayes factor ($B_{m}$) in favor of the more complex model.
Adopted from \citet{trotta_bayes_2008}. A Bayes factor of 150 can
be considered a ``strong'' detection, corresponding to approximately
3.6$\sigma$ significance in the frequentist's framework.\label{tab:Translation-table-between}}
\end{table}

\subsection{Nested Sampling for Atmospheric Retrieval\label{sub:Nested-Sampling-for}}

We employ the multimodal nested sampling algorithm (MultiNest) to
efficiently compute the Bayesian evidences of alternative retrieval
models \citep{skilling_nested_2004,feroz_multimodal_2008,feroz_multinest:_2009}.
The joint posterior probability distribution of the atmospheric parameters
for a given retrieval model is obtained as a by-product. The MultiNest
algorithm was developed as a Bayesian inference tool for cosmology
and particle physics, and we find that it provides substantial advantages
for atmospheric retrieval over techniques based on Markov Chain Monte
Carlo (MCMC).

\subsubsection{Nested Sampling versus MCMC}

Parameter estimations in many astrophysical contexts are typically
performed using the MCMC technique with the Metropolis-Hastings algorithm
or its variants such as the Gibbs sampler. The two main disadvantages
of the MCMC techniques are, however, that MCMC does not directly provide
the Bayesian evidence for comparing retrieval models of different
complexities, and that MCMC becomes inefficient when sampling from
a posterior distribution with multiple separate modes or elongated
curving degeneracies. In \citet{benneke_atmospheric_2012}, we were
able to search for separated modes in the posterior distribution using
the parallel tempering MCMC technique. Calculating the Bayesian evidences
for Bayesian model comparisons would, however, require a second step,
such as a Restricted Monte Carlo (RMC) integration. Restricted Monte
Carlo (RMC) becomes extremely inefficient, however, when the joint
posterior distribution shows strongly curved correlations, such as
those encountered in atmospheric retrieval problems.

\subsubsection{The Nested Sampling Algorithm}

Nested sampling is an alternative Monte Carlo technique for Bayesian
inference. It enables one to efficiently compute the Bayesian evidence
and provides the posterior distribution for parameter estimations
as a by-product. A full discussion is provided in \citep{skilling_nested_2004,feroz_multimodal_2008}.
Here, we provide a brief description of the main ideas. For clarity,
we do not explicitly state the model $M_{i}$ and the data $\boldsymbol{D}$
because the calculation of the Bayesian evidence is conducted for
each model and data set individually.

The basic idea behind nested sampling is to transform the multi-dimensional
integral for the computation of the Bayesian evidence (Equation (\ref{eq:BayesianEvidence}))
into the one-dimensional integral

$ $
\begin{equation}
\mathcal{Z}=\int_{0}^{1}\mathcal{L}^{*}\left(V\right)dV.\label{eq:EvidenceNested}
\end{equation}

The integration variable $V$ in Equation (\ref{eq:EvidenceNested})
is the ``prior volume'' defined as

\begin{equation}
V\left(\mathcal{L}^{*}\right)=\int_{\mathcal{L}\ensuremath{\left(\boldsymbol{\theta}\right)}>\mathcal{L}^{*}}\pi\text{\ensuremath{\left(\boldsymbol{\boldsymbol{\theta}}\right)}}\mathrm{d}^{N}\boldsymbol{\theta}.\label{eq:PriorVolume}
\end{equation}

The prior volume $V\left(\mathcal{L}^{*}\right)$ is the prior probability
density integrated over all regions in the parameter space for which
the likelihood function $\mathcal{L}\ensuremath{\left(\boldsymbol{D}|\boldsymbol{\theta}\right)}$
exceeds the value $\mathcal{L}^{*}$. It is a monotonically decreasing
function of the likelihood limit $\mathcal{L}^{*}$ because the volume
in the parameter space that meets the criteria $\mathcal{L}\ensuremath{\left(\boldsymbol{\theta}\right)}>\mathcal{L}^{*}$
decreases as the likelihood limit $\mathcal{L}^{*}$ is increases.
The extreme values are $V=1$ for $\mathcal{L}^{*}=0$, in which case
the integration is performed over the entire prior parameter space,
and $V=0$ for $\mathcal{L}^{*}>\mathrm{max\left(\mathcal{L}\ensuremath{\left(\boldsymbol{D}|\boldsymbol{\theta}\right)}\right)}$.
The function $\mathcal{L}^{*}\left(V\right)$ in Equation \ref{eq:EvidenceNested}
is the inverse function of the prior volume $V\left(\mathcal{L}^{*}\right)$
and is thus also monotonically decreasing. 

\begin{figure}[t]
\includegraphics[width=1\columnwidth]{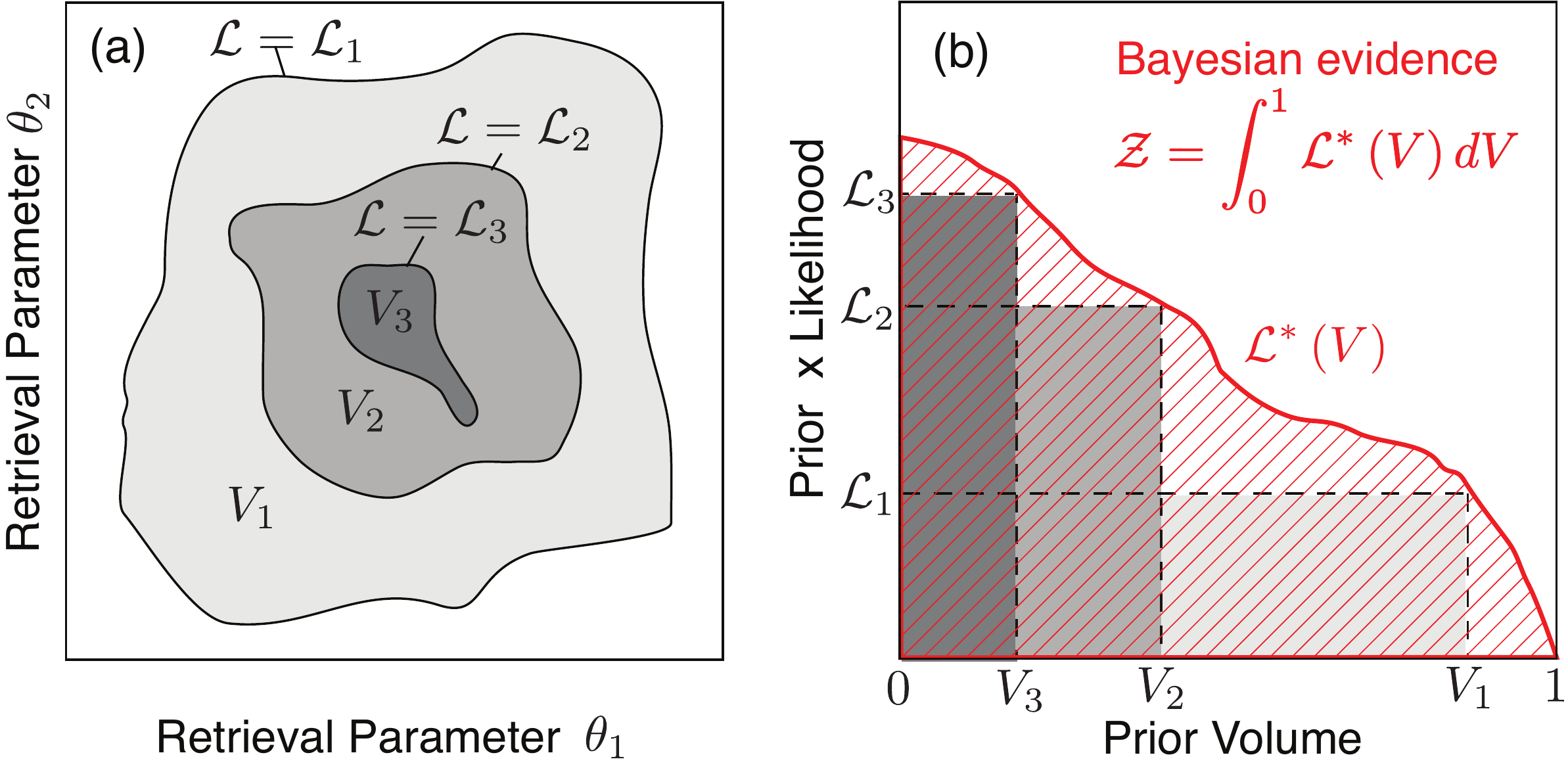}

\caption{Graphical interpretation of the prior volume and Bayesian evidence
for a two dimensional parameter space. The prior volume $V_{i}=V\left(\mathcal{L}_{i}\right)$
is the prior probability density integrated over all regions in the
parameter space for which the likelihood function $\mathcal{L}\ensuremath{\left(\theta_{1},\theta_{2}\right)}$
exceeds the value $\mathcal{L}_{i}$ . For a two-dimensional parameter
space with a uniform prior, the prior volume is the projected area
of the region in the parameter space for which $\mathcal{L}\ensuremath{\left(\theta_{1},\theta_{2}\right)}$
exceeds the value $\mathcal{L}_{i}$ (panel (a)). The Bayesian evidence
$\mathcal{Z}$ can be regarded as the geometric volume between the
$\theta_{1}$,$\theta_{2}$-plane and the likelihood surface $\mathcal{L}\ensuremath{\left(\theta_{1},\theta_{2}\right)}$.
Once a sequence of $\mathcal{L}_{i},V_{i}$ pairs is known, the Bayesian
evidence can be approximated using standard quadrature methods (panel
(b))$ $.\label{fig:In-the-two}}
\end{figure}

Figure~\ref{fig:In-the-two} provides a graphical interpretation
of the prior volume $V$ and equation~(\ref{eq:EvidenceNested})
for a two-dimensional parameter space. For two dimensional parameter
spaces with the uniform prior probability $\pi(\theta_{1},\theta_{2})=\pi_{\mathrm{uni}}$,
the Bayesian evidence can be regarded as the geometric volume between
the $\theta_{1}$,$\theta_{2}$-plane and the surface of the likelihood
function $\mathcal{L}\ensuremath{\left(\boldsymbol{D}|\theta_{1},\theta_{2}\right)}$,
multiplied by $ $$\pi_{\mathrm{uni}}$ (see equation (\ref{eq:BayesianEvidence})).
The prior volume $V\left(\mathcal{L}^{*}\right)$ can be interpreted
as the area within iso-likelihood contour $\ensuremath{\mathcal{L}\ensuremath{\left(\boldsymbol{D}|\theta_{1},\theta_{2}\right)}=\mathcal{L}^{*}}$
projected onto the $\theta_{1},\theta_{2}$ -plane, again multiplied
by uniform prior $\pi_{\mathrm{uni}}$. 

The Bayesian evidence in Equation~\ref{eq:EvidenceNested} can be
approximated using standard quadrature methods if the inverse function
$\mathcal{L}^{*}\left(V\right)$ of Equation (\ref{eq:PriorVolume})
can be evaluated at a sequence of values $0\leqslant V_{M}<\cdots<V_{i}<\cdots<V_{1}\leqslant1$.
Using the trapezium rule, we obtain 

\begin{equation}
\mathcal{Z}=\sum_{i=1}^{M}w_{i}\mathcal{L}_{i}.\label{eq:}
\end{equation}

where $\mathcal{L}_{i}$ is the likelihood limit $\mathcal{L}^{*}\left(V_{i}\right)$
corresponding to $V_{i}$, the weights $w_{i}$ are given by $w_{i}=\frac{1}{2}\left(V_{i-1}-V_{i+1}\right)$,
and $M$ is the number of values in the sequence.

In nested sampling, a sequence of $\mathcal{L}^{*}\left(V_{i}\right)$
values is generated as follows. The algorithm is initialized by randomly
drawing a user-specified number of ``active samples'' ($N\approx50\ldots10000$)
from the full prior probability distribution $\pi\text{\ensuremath{\left(\boldsymbol{\boldsymbol{\theta}}\right)}}$
and computing their likelihood $\mathcal{L}\ensuremath{\left(\boldsymbol{\theta}\right)}$
according to Section~\ref{eq:likelihoodfct}. These samples are distributed
randomly across the full prior parameter space and the prior volume
$V_{0}=V\left(\mathcal{L}^{*}=0\right)$ is 1. The first iteration
is started by sorting the samples in order of their likelihood. The
sample with the lowest likelihood $\mathcal{L}_{1}$, i.e. the sample
with the worst fit to the data, is then removed and replaced by a
new sample. The new sample is again drawn from the prior distribution,
but subject to the constraint that the likelihood of the new sample
is higher than $\mathcal{L}_{1}$. If necessary, the drawing of the
new sample is repeated until a sample is found with $\mathcal{L}\ensuremath{\left(\boldsymbol{\theta}\right)}>\mathcal{L}_{1}$.
After the replacement, all active samples meet the condition $\mathcal{L}\ensuremath{\left(\boldsymbol{\theta}\right)}>\mathcal{L}_{1}$
and they are uniformly distributed within the iso-likelihood contour
$\mathcal{L}^{*}=\mathcal{L}_{1}$. The new prior volume $V_{1}$
contained within the iso-likelihood contour $\mathcal{L}^{*}=\mathcal{L}_{1}$
can be written as $V_{1}=t_{1}V_{0}$, where the shrinkage ratio $t_{1}$
is a random number smaller than 1. The probability distribution of
the shrinkage, $p\left(t_{1}\right)$, is the distribution for the
largest of $n$ samples drawn from a uniform distribution between
0 and 1. The distribution $p\left(t_{1}\right)$ describes the decrease
in prior volume between subsequent iso-likelihood contours $\mathcal{L}^{*}=V_{0}$
and $\mathcal{L}^{*}=V_{1}$ in a probabilistic way because the active
samples are uniformly distributed within the iso-likelihood contours.

In subsequent iterations, the process of replacing the sample with
the lowest likelihood $\mathcal{L}_{i}$ is repeated, and the corresponding
prior volume $V\left(\mathcal{L}_{i}\right)$ is repeatedly tightened.
The nested sampling algorithm thus progresses through nested shells
of iso-likelihood contours with continually decreasing prior volumes,
until the regions of the highest likelihoods are localized. The prior
volume $\mathrm{V_{i}}$ after iteration $i$ can be approximated
as

\begin{equation}
V_{i}=\exp\left(-i/N\right).\label{eq:tightening}
\end{equation}

Equation (\ref{eq:tightening}) approximate the prior volume at iteration
$i$ because each value of $t_{i}$ is independent and the mean and
standard deviations of $\log\left(t\right)$ are $\mathrm{E}\left[\log\left(t\right)\right]=-1/N$
and $\sigma\left[\log\left(t\right)\right]=1/N$, resulting in $\log\left(V_{i}\right)\approx\left(i\pm\sqrt{i}\right)/N$.

\subsubsection{Simultaneous Ellipsoidal Nested Sampling}

The main challenge in efficiently computing the Bayesian evidence
is to efficiently generate random samples within the iso-likelihood
contour $\mathcal{L}^{*}=\mathcal{L}_{i}$. If one continued to draw
random samples from the full prior probability distribution $\pi\text{\ensuremath{\left(\boldsymbol{\boldsymbol{\theta}}\right)}}$
as in the first iteration, the chance of randomly finding one that
is within the iso-likelihood contour $\mathcal{L}^{*}=\mathcal{L}_{i}$
would decrease exponentially as the likelihood limit $\mathcal{L}_{i}$
is raised.

The algorithm employed in this work uses the simultaneous ellipsoidal
nested sampling method developed in \citet{feroz_multinest:_2009}.
At each iteration, the full set of active samples is partitioned according
to local clustering and an optimum number of ellipsoids are constructed
such that the union of the ellipsoidal volumes tightly encompasses
all samples. At early iterations, a small number of large ellipsoids
cover almost the entire prior parameter space. As the active sample
become confined to the regions within tighter iso-likelihood contours,
the ellipsoids encompass only the region(s) of high likelihood $\mathcal{L}\ensuremath{\left(\boldsymbol{\theta}\right)}$.
New random samples are drawn from within the ellipsoids, thus the
random samples have a high likelihood $\mathcal{L}\ensuremath{\left(\boldsymbol{\theta}\right)}$
with higher probability, resulting in a higher sample acceptance rate
and higher efficiency of the nested sampling algorithms. The algorithm
is robust to multimodal posterior distributions and elongated curving
degeneracies, while remaining efficient for simpler distributions.
Refer to \citet{feroz_multinest:_2009} for a detailed description
of the algorithm employed for partitioning and construction of the
ellipsoids.

\subsubsection{Convergence and Numerical Uncertainty}

The iterations are continued until the Bayesian evidence is determined
to a specified precision $\Delta\mathcal{Z}$. In this work, we terminated
the iterations once the logarithm of the evidence did not change by
more than $\Delta\left(\log\left(\mathcal{Z}\right)\right)=0.001$.
The final uncertainty can be estimated following \citet{sivia_data_2006}
as

\begin{equation}
\log\left(\mathcal{Z}\right)\approx\log\left(\sum_{i=1}^{M}w_{i}\mathcal{L}_{i}\right)\pm\sqrt{\frac{\mathcal{H}}{N}},
\end{equation}

where $\mathcal{H}$ is the \textit{information, }or\textit{ negative
entropy, }

\begin{equation}
\mathcal{H}\approx\sum_{i=1}^{M}\frac{w_{i}\mathcal{L}_{i}}{\mathcal{Z}}\log\left(\frac{\mathcal{L}_{i}}{\mathcal{Z}}\right).
\end{equation}

\subsubsection{Parameter Estimation}

Once the Bayesian evidence $\mathcal{Z}$ is determined, the joint
posterior distribution of the atmospheric parameters can be constructed
using all active and discarded samples that were generated during
the nested sampling iterations. Each sample is assigned a weight according
to 
\begin{equation}
W_{i}=\frac{w_{i}\mathcal{L}_{i}}{\mathcal{Z}}.
\end{equation}

The weighted samples can be used to plot the posterior distributions
and calculate the statistical measures such as the mean and covariances
matrix.

\section{Results: Distinguishing between $\mathrm{H_{2}}$-dominated and $\mathrm{H_{2}O}$-dominated
atmospheres\label{sec:Results---General}}

In this Section, we demonstrate that a promising strategy to distinguish
between cloudy hydrogen-dominated atmospheres and clear water- or
volatile-dominated atmospheres is to observe individual water absorption
features in the planet's transmission spectrum at moderate spectral
resolution ($R\thickapprox100$). The unambiguous distinction is possible
based on the effect of the mean molecular mass on the wing slopes
of the water features and the relative depths of different water absorption
features. As a case study, we investigate, for the super-Earth GJ~1214b,
how much the observational uncertainty needs to be improved compared
to published transmission spectra to reliably detect NIR water absorption
features and to unambiguously distinguish between the hydrogen-dominated
and water-dominated scenarios. All results derived in this Section
can be generalized to other super-Earths and Neptunes such as HD~97658b,
55~Cnc~e, and GJ~436b using the scalings provided in Section \ref{sub:Scaling-Law-for}
and \ref{sub:HD97658b,-55Cnce,-and}.

\subsection{Distinction Based on NIR Transmission Spectroscopy \label{sub:Distinguishing-between-Water-Ric}}

Our main finding is that an unambiguous distinction will be possible
for a GJ~1214b-like planet if the observational uncertainty in the
transit depth measurements can be reduced by a factor of $\sim3$
compared with the published \textit{VLT} observations by \citet{bean_ground-based_2010}
and \textit{HST WFC3} observations by \citet{berta_flat_2012}. Alternatively,
the distinction can be achieved using only \textit{HST WFC3} observations
if the uncertainty can be reduced by a factor of $\sim5$ compared
to the previous result by \citet{berta_flat_2012} (Section \ref{sub:Distinction-Using-HST}). 

\begin{figure}[p]
\selectlanguage{british}%
\noindent \begin{centering}
\includegraphics[scale=0.32]{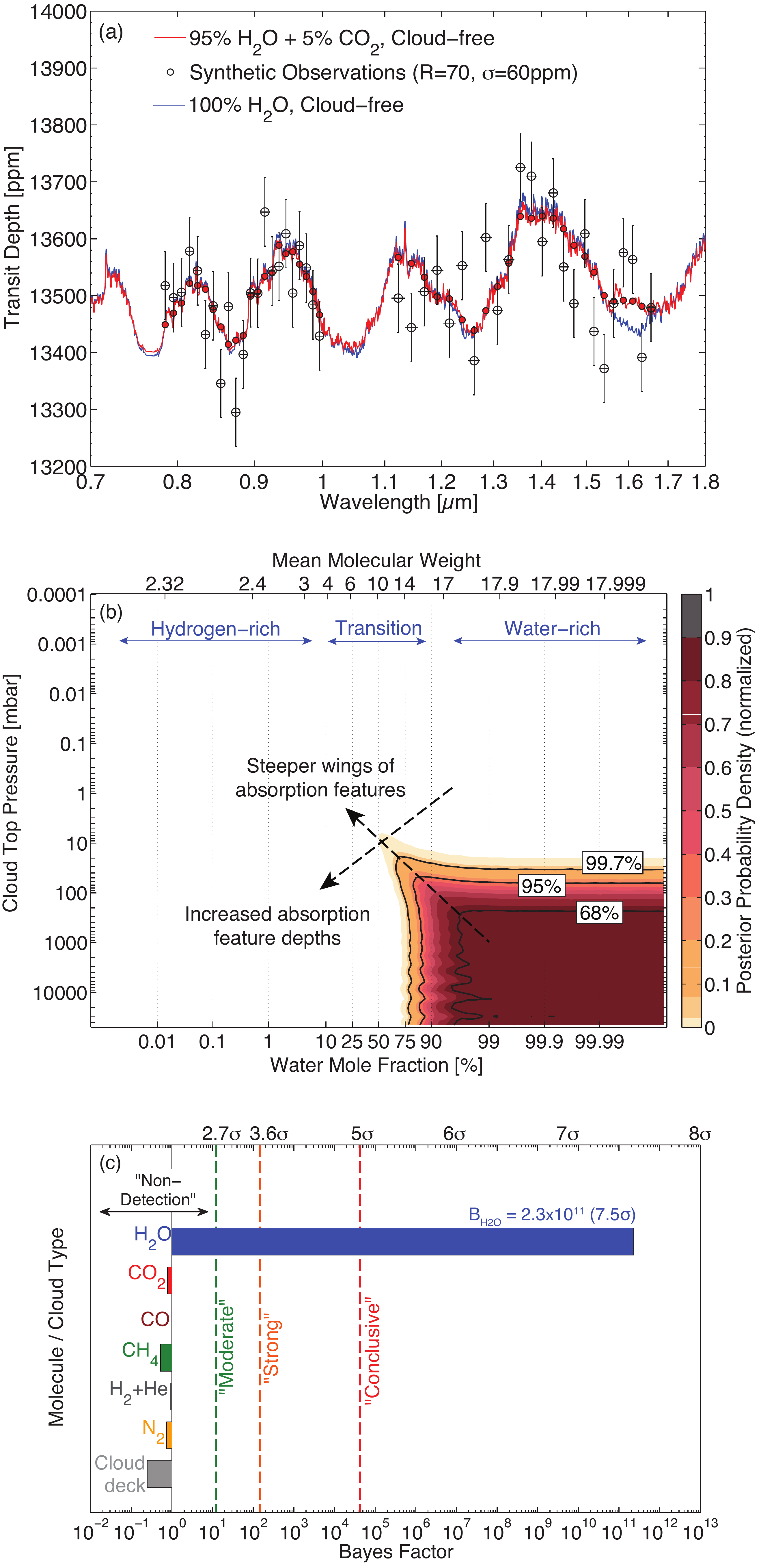}
\par\end{centering}

\selectlanguage{english}%
\caption{Synthetic 60~ppm observations and retrieval results for a cloud-free,
water-dominated atmosphere on GJ~1214b. Panel (a) shows synthetic
observations of a 95\% $\mathrm{H_{2}O}$ + 5\% $\mathrm{CO_{2}}$
atmosphere covering the spectral range of the \textit{VLT }and \textit{HST
WFC3} observations by \citet{bean_ground-based_2010} and \citet{berta_flat_2012}.
Panel (b) illustrates the posterior probability distribution derived
from the synthetic observations as a function of cloud top pressure
and mean molecular weight for a simple two-component atmosphere composed
of hydrogen and water vapor. The black contour lines indicate the
68\%, 95\%, and 99.7\% Bayesian credible regions. The colored shading
illustrates the regions of high posterior probability density. The
horizontal axis is scaled as the centered-log-ratio parameter $\xi_{1}=\log\left(X_{H2O}/\sqrt{X_{H2O}\cdot X_{H2}}\right)$.
60~ppm observations would constrain the mean molecular mass to $\mu_{\mathrm{atm}}>13$
at 99.7\% probability ($3\sigma$). Such observations would conclusively
rule out a hydrogen-rich nature of the atmosphere. Panel (c) illustrates
the Bayes factors describing the detection confidences of molecular
gases and clouds based on the synthetic observations. 60~ppm observations
would enable a robust $B_{\mathrm{H_{2}O}}=2\cdot10^{12}\,\left(7.8\sigma\right)$
detection of water absorption. The 5\% $\mathrm{CO_{2}}$ remain undetectable
because strong water absorption blocks the $\mathrm{CO_{2}}$ signature.\label{fig:H2O_Distinguish}}
\end{figure}

\begin{figure}[p]
\selectlanguage{british}%
\noindent \begin{centering}
\includegraphics[scale=0.32]{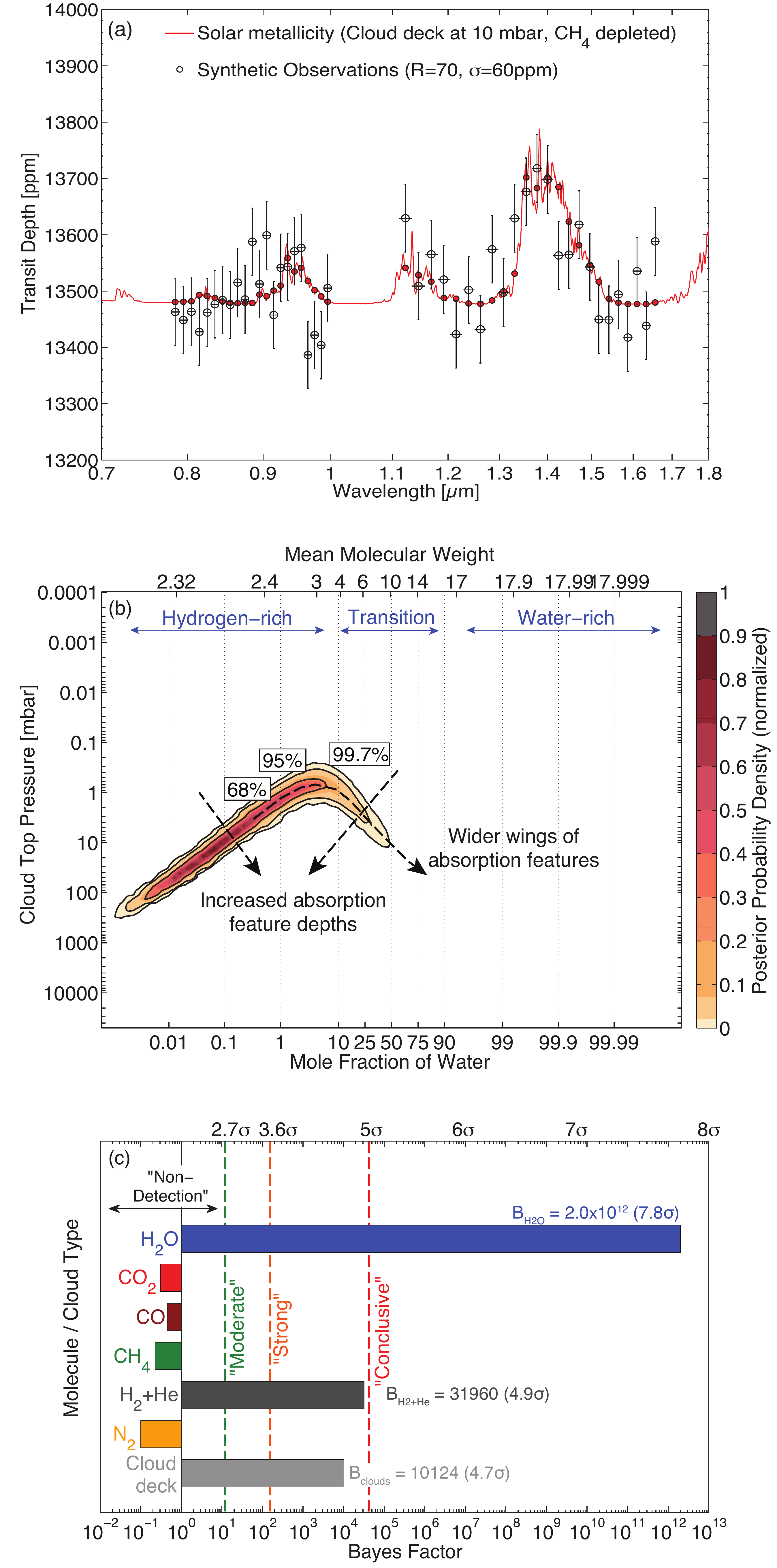}
\par\end{centering}

\selectlanguage{english}%
\caption{Synthetic 60~ppm observations and retrieval results for a $\mathrm{H_{2}}$-dominated
atmosphere with high-altitude clouds on GJ~1214b. Panel (a) shows
synthetic observations of a methane-depleted, solar metallicity atmosphere
with thick gray clouds extending up to the 10~mbar level. The spectral
range is equivalent to the \textit{VLT }and \textit{HST WFC3} observations
by \citet{bean_ground-based_2010} and \citet{berta_flat_2012}. The
methane depletion and vertical extent of the clouds were chosen to
ensure that the distinction between the $\mathrm{H_{2}O}$-dominated
and $\mathrm{H_{2}}$-dominated scenarios is independent of other
molecular absorbers. Panels (b) and (c) display the retrieval results
as explained in Figure \ref{fig:H2O_Distinguish}. 60~ppm observations
would be able to confirm the hydrogen-dominated nature by setting
an $3\sigma$ upper limit on mean molecular mass at $\mu_{\mathrm{ave}}=10$.
The observations would be sufficient to robustly infer the presence
of water and hydrogen, and clouds at high confidence.\label{fig:H2_Distinguish}}
\end{figure}

\begin{figure}[H]
\selectlanguage{british}%
\noindent \begin{centering}
\includegraphics[scale=0.32]{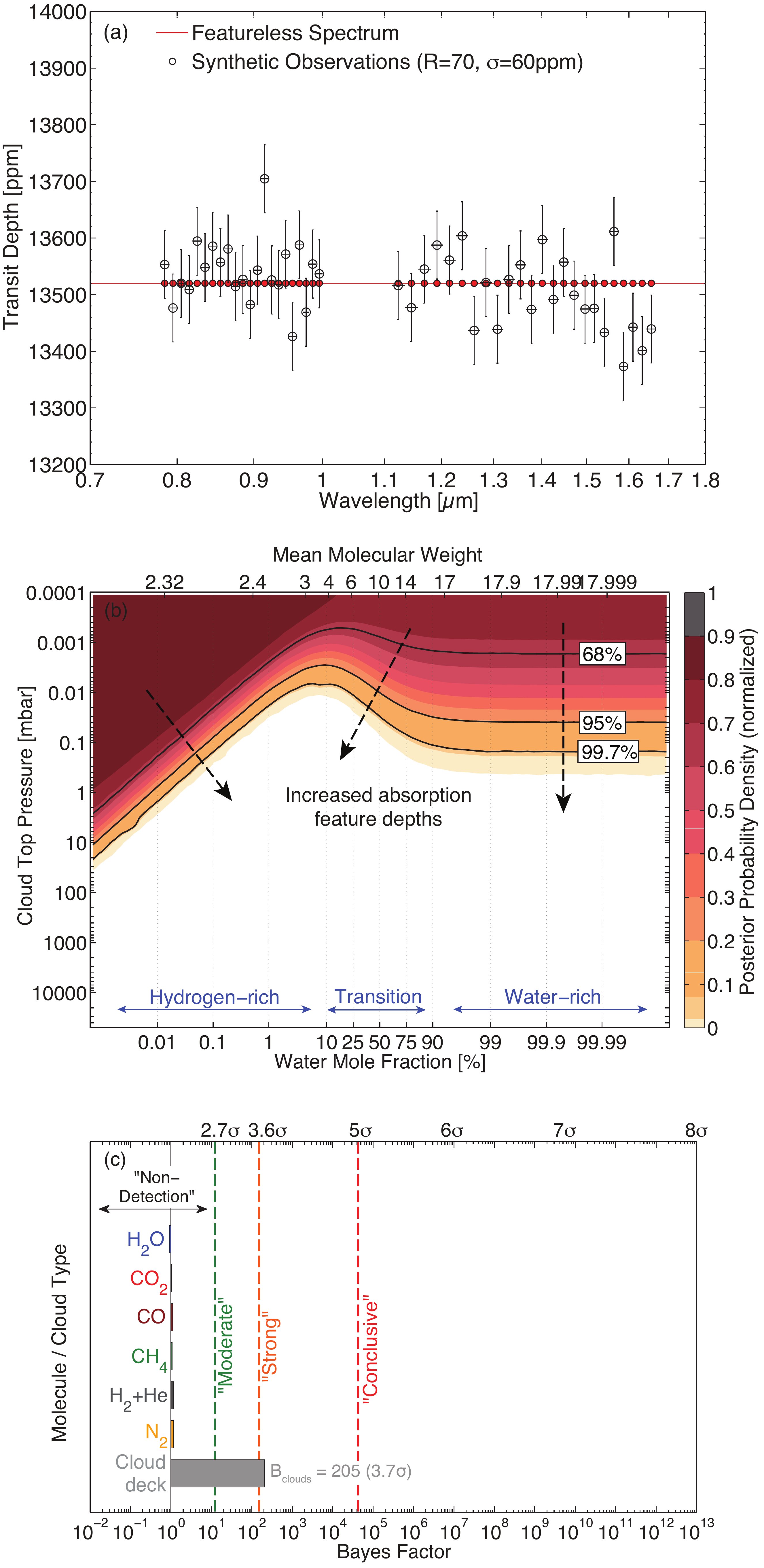}
\par\end{centering}

\selectlanguage{english}%
\caption{Synthetic 60~ppm observations and retrieval results for a featureless
spectrum of GJ~1214b. Panel (a) displays synthetic observations GJ~1214b
as it will appear if a thick cloud deck surrounds the planet at above
the $0.1\,\mathrm{mbar}$ pressure level or no atmosphere is present.
Panels (b) and (c) display the retrieval results derived from synthetic
observations as explained in the caption of Figure \ref{fig:H2O_Distinguish}.
At 60~ppm, a flat spectrum would rule out thick, cloud-free scenarios
for all water mole fractions between the one expected for solar composition
($X_{\mathrm{H_{2}O}}\sim0.04\%$) and water-dominated ($X_{\mathrm{H_{2}O}}\rightarrow100\%$).
At $3\sigma$, water-dominated scenarios are possible only if a cloud
deck is present above the $0.1\,\mathrm{mbar}$ pressure level. Hydrogen-dominated
scenarios are compatible with the flat spectrum at 60~ppm only if
the partial pressure of $\mathrm{H_{2}O}$ is below $\unit[2]{\mu bar}$
at the uppermost cloud deck surface. Panel (c) shows that no molecular
detections can be inferred from the featureless spectrum. Clouds are
likely but cannot conclusively be inferred because atmospheric scenarios
without NIR absorbers could theoretically match the featureless spectrum.
\label{fig:Flat_Distinguish}}
\end{figure}

For the following discussions, we consider synthetic observations
of three alternative scenarios for the atmosphere of GJ~1214b. The
scenarios are a cloud-free, water-dominated atmosphere composed of
95\% $\mathrm{H_{2}O}$ and $\mathrm{CO_{2}}$ (Figure \ref{fig:H2O_Distinguish}),
a hydrogen-dominated atmosphere with solar metallicity ($\sim400$~ppm
$\mathrm{H_{2}O}$) and high-altitude clouds (Figure \ref{fig:H2_Distinguish}),
and flat transmission spectra due to the lack of atmospheres or the
presence of an extremely high cloud deck (Figure \ref{fig:Flat_Distinguish}).
The hydrogen-dominated scenario is depleted in methane and clouds
are present at 10~mbar such that depths of the water absorption resemble
the ones of the cloud-free water-dominated scenario. These choices
ensure that the distinction between the water-dominated and hydrogen-dominated
scenarios is based the effect of the mean molecular mass on relative
sizes and shapes of the water absorption features and independent
of the abundance or the presence of other molecular species.

We find that spectral observations covering the spectral ranges $\unit[0.78-1]{\mu m}$
\citep{bean_ground-based_2010} and $\unit[1.1-1.8]{\mu m}$ \citep{berta_flat_2012},
can unambiguously distinguish between water-dominated atmospheres
and hydrogen-dominated atmospheres with water absorption if the observational
uncertainty can be reduced to 60~ppm at a spectral resolving power
of $R=70$ ((Figure \ref{fig:H2O_Distinguish}-\ref{fig:Flat_Distinguish}).
This corresponds to an improvement of $\sim3$ compared with the published
uncertainties of $180-200\,\mathrm{ppm}$ by \citet{bean_ground-based_2010}
and \citet{berta_flat_2012}.

\paragraph{(a) Water-dominated scenario.}

If GJ~1214b is surrounded by a cloud-free water-dominated atmosphere
(Figure \ref{fig:H2O_Distinguish}), transit depth measurements with
60~ppm uncertainty will constrain the mean molecular mass to $\mu_{\mathrm{atm}}>13$
at 99.7\% probability ($3\sigma$). Such observations would conclusively
rule out a hydrogen-rich nature of GJ~1214b. Taking a simple two-component
model atmosphere composed of water vapor and hydrogen gas, we could
infer at $3\sigma$ that $>70\%$ of the atmosphere would need to
be water vapor, leaving only a maximum of $30\%$ for hydrogen-helium
gas (Figure \ref{fig:H2O_Distinguish}(b)). From 60~ppm observations,
we would also infer at $3\sigma$ that the atmosphere is cloud-free
down to at least the $20$~mbar level. The posterior probability
density would be maximum for water fractions above 90\% and cloud
top pressures above 100~mbar. Changes of the water fractions above
90\% have little effect on the water absorption features because the
mean molecular mass remains largely unchanged. An increase in cloud
top pressure above $\unit[100]{mbar}$, similarly, has negligible
effects on the observable spectrum because the high opacity of water
vapor across the full NIR spectrum prevents probing deeper atmospheric
layers through transmission spectroscopy.

\paragraph{(b) Hydrogen-rich scenario with high-altitude clouds.}

If, alternatively, GJ~1214b is surrounded by a hydrogen gas envelope
(Figure \ref{fig:H2_Distinguish}), transit depth measurements with
60~ppm uncertainty will be able to confirm the hydrogen-dominated
nature of the atmosphere, even if the depth of water absorption features
resembles the one of a water-dominated scenario. 60~ppm transit observations
would provide sufficient information on the relative strengths of
the absorption features and the steepnesses of the feature wings to
infer a mean molecular mass below 10 at $3\sigma$ (Figure \ref{fig:H2_Distinguish}(b)).
A mean molecular mass below 10 would mean that hydrogen/helium gas
makes up at least 50\% of the atmosphere. The observations would also
confirm the presence of an upper cloud deck with a cloud-top pressure
between 0.2~mbar and 200~mbar at 99.7\% confidence. 

Stronger independent constraints on the water mole fraction in the
hydrogen-dominated atmosphere would not be available despite a $7\sigma$
detection of water absorption because a strong correlation exists
between the water mole fraction and the cloud-top pressure for the
transmission spectra of a hydrogen-dominated atmosphere. The strong
correlation is present because atmospheric scenarios with different
combinations of mean molecular masses and cloud-top pressures can
lead to identical feature depths. 

It is interesting to note that even though hydrogen/helium gas shows
no direct absorption features, the presence of hydrogen/helium gas
in the atmosphere can be inferred from the NIR observations if hydrogen/helium
gas is present in sufficient amounts to affect the mean molecular
mass (Figure \ref{fig:H2_Distinguish}(c)). The reason is that no
other plausible atmospheric component can explain the mean molecular
mass below that of water ($\mu=18$) and methane ($\mu=16$) in the
temperature range expected for GJ1214b. Clouds and hazes can be inferred
from transmission spectra if present at high altitude because they
lower the transit depth variations below that expected for a clear
atmosphere of a given atmospheric composition.

\paragraph{(c) Featureless spectrum.}

A planet's transmission spectrum may be featureless if the planet
lacks a gaseous atmosphere or clouds are present at high altitude.
If a featureless spectrum with an observational uncertainty of 60~ppm
was observed for GJ~1214b (Figure \ref{fig:Flat_Distinguish}), the
observations would be sufficient to rule out the presence of cloud-free,
water-rich atmospheres at high significance (Figure \ref{fig:H2O_Distinguish}(e-f)).
For the full range of water mole fractions between solar composition
($X_{\mathrm{H_{2}O}}\sim0.04\%$) and water-dominated ($X_{\mathrm{H_{2}O}}\rightarrow100\%$),
we could conclude that cloud top pressures must be below 1~mbar.
The observations would additionally rule out scenarios with cloud-top
pressures down to 0.01 mbar for water mole fractions between 1\% and
10\%.

As a consistency check, Figure \ref{fig:Flat_Distinguish}(c) shows
that no molecular species can be inferred from the flat spectrum.
It is interesting to note that the Bayes factor for the presence of
clouds is not conclusive despite the flat spectrum. The reason for
this is that high altitude clouds are only one explanation for measuring
a flat spectrum. For example, the absence of strong absorber such
as water and methane in combination with a high mean molecular mass
would similarly explain the lack of any features at the 60~ppm level.
A comparison between Figures \ref{fig:Flat_Distinguish}(c) and \ref{fig:H2_Distinguish}(c),
in fact, shows that the presence of weak molecular features may lead
to a higher confidence in the presence of clouds than a flat spectrum.
Absorption features enable us to infer the composition and the depths
and shapes of the features then enable us to determine whether clouds
must be present or not. This information is not available if a flat
spectrum is observed.

\subsection{Distinction Based Solely on \textit{HST WFC3} Observations\label{sub:Distinction-Using-HST}}

\begin{figure}[t]
\selectlanguage{british}%
\includegraphics[width=1\columnwidth]{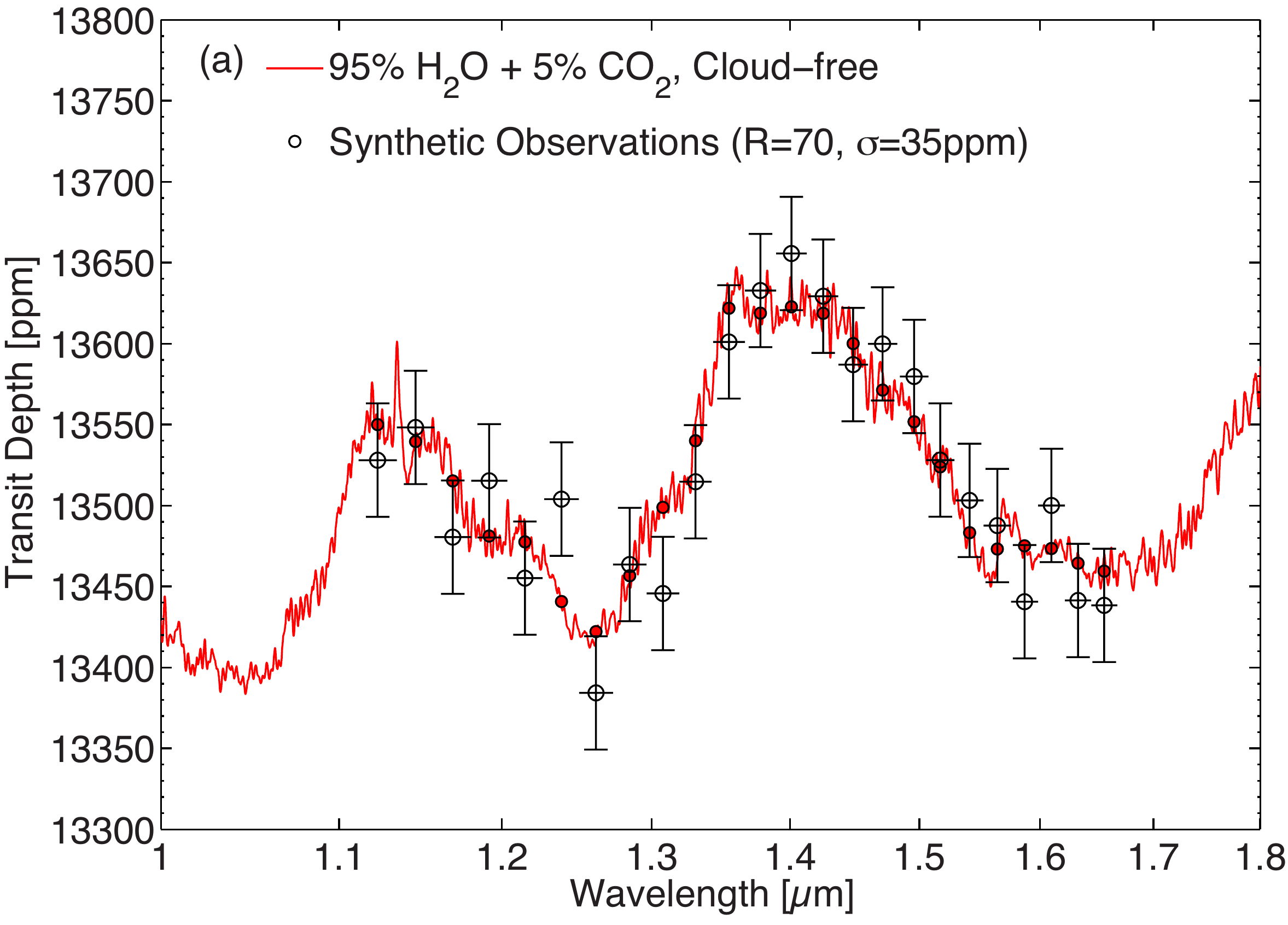}

\includegraphics[width=1\columnwidth]{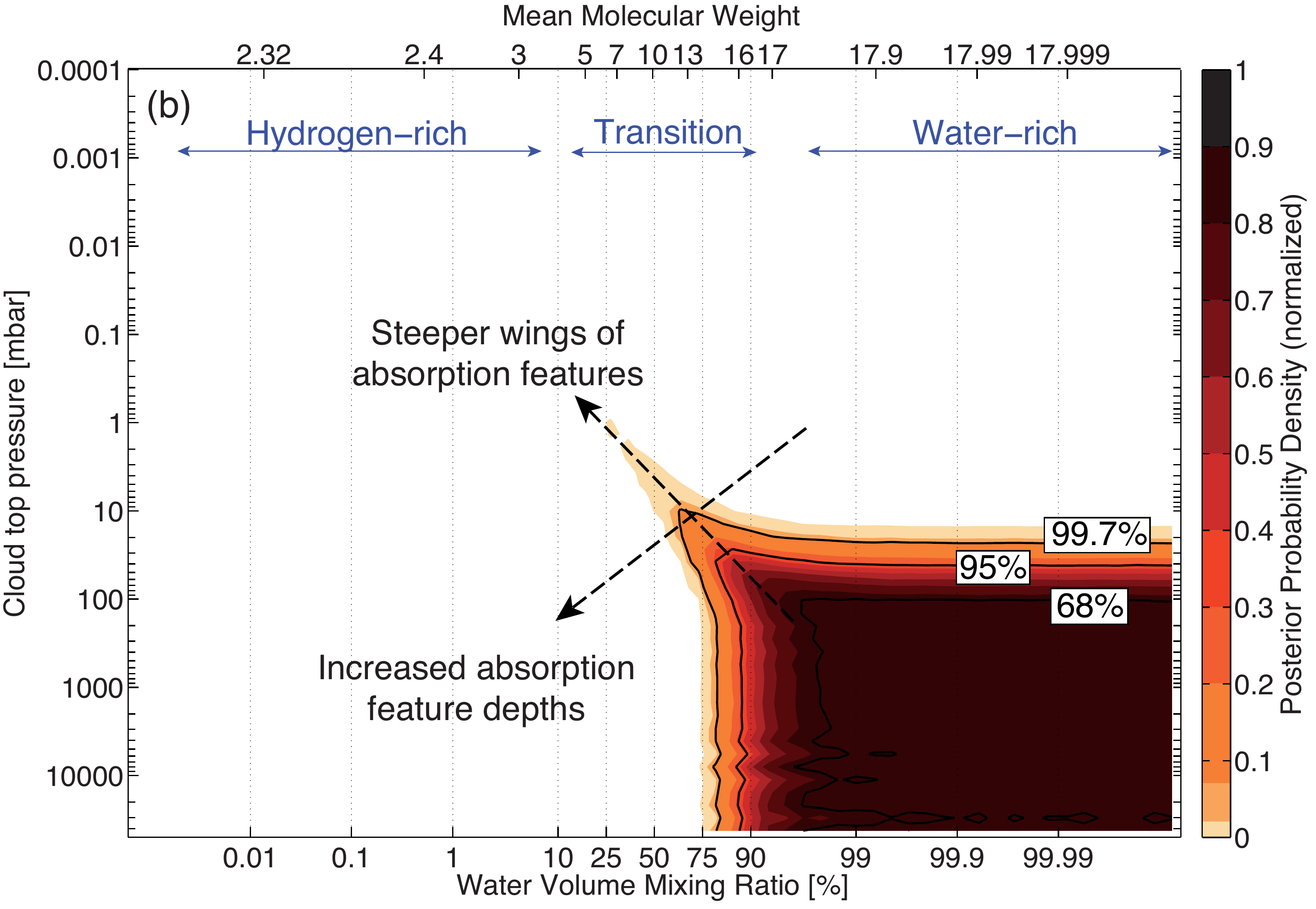}

\selectlanguage{english}%
\caption{Synthetic 35~ppm observations and retrieval results for a cloud-free,
$\mathrm{H_{2}O}$-dominated atmosphere on GJ~1214b. The atmospheric
scenario ( 95\% $\mathrm{H_{2}O}$ + 5\% $\mathrm{CO_{2}}$) is identical
to the one in Figure \ref{fig:H2O_Distinguish}, however, the synthetic
observations cover only the spectral range of \textit{HST WFC3.} Panel
(b) illustrates the posterior probability distribution as a function
of cloud top pressure and mean molecular weight as explained in Figure
\ref{fig:H2O_Distinguish}. 35~ppm observations in the spectral range
of \textit{HST WFC3} ($1.1-1.7\,\mathrm{\mu m}$) would constrain
the mean molecular mass to $\mu_{\mathrm{atm}}>11$ at 99.7\% probability
($3\sigma$). Such observations would conclusively rule out a hydrogen-rich
nature of the atmosphere. Smaller observational uncertainties are
required compared to Figure \ref{fig:H2O_Distinguish} because the
weaker water absorption bands at $0.8-1\,\mathrm{\mu m}$ are not
captured by \textit{HST WFC3} observations alone. \label{fig:HSTonly}.}
\end{figure}

The distinction between hydrogen-dominated and water-dominated scenarios
is possible from \textit{HST WFC3} observations if the observational
uncertainty can be reduced to below $\sim35$~ppm (Figure \ref{fig:HSTonly}).
The main advantage of augmenting the \textit{HST WFC3} observation
with observations at $0.8-1\,\mathrm{\mu m}$ (such as the ones provided
by $VLT$) is that $0.8-1\,\mathrm{\mu m}$ contains weaker absorption
features that, together with the strong absorption feature at 1.38~$\mathrm{\mu m}$,
enable a good comparison between the relative depths of the strong
and weak water absorption features. These relative depths, in turn,
constrain the mean molecular mass and thus enable the distinction
between hydrogen-dominated atmospheres and those dominated by water
or other volatiles. If only \textit{HST WFC3} observations are available,
the observational uncertainty needs to be sufficient to capture the
feature wings between $\unit[1.1-1.7]{\mu m}$ at higher precision
because the information from the relative depths of the features will
not be available (Figure \ref{fig:HSTonly}(a)).

\subsection{Detectability of NIR Water Absorption \label{sub:Weak-Constraints-on}\label{sub:Detectability-of-NIR}}

It is important to recognize that detections of water absorption are
a necessary, but not sufficient condition to determine the water fraction
and nature of low-density super-Earths. Section \ref{sub:Distinguishing-between-Water-Ric}
showed that transit observations with $\lesssim50$~ppm observational
uncertainty are required to distinguish hydrogen-dominated and water-dominated
scenarios. The presence of water absorption in a water-dominated atmosphere
can already be detected at high confidence for transit observations
with up to 80~ppm uncertainty. 80~ppm observations at $R=70$ would
provide a $B_{\mathrm{H_{2}O}}=40000\,\left(5\sigma\right)$ detection
of water absorption and would only require an improvement by a factor
of 2.5 compared to \citet{bean_ground-based_2010} and \citet{berta_flat_2012}. 

\begin{table*}[t]
\noindent \begin{centering}
{\scriptsize }%
\begin{tabular}{llccc}
\hline 
{\scriptsize Retrieval model} & {\scriptsize Retrieval model parameters} & {\scriptsize Evidence } & {\scriptsize Best-fit} & {\scriptsize Bayes factor }\tabularnewline
 &  & {\scriptsize $\ln\left(\mathcal{Z}_{i}\right)$} & {\scriptsize $\chi_{\mathrm{best-fit}}^{2}$} & {\scriptsize $B_{i}=\mathcal{Z}_{0}/\mathcal{Z}_{i}$}\tabularnewline
\hline 
\multirow{1}{*}{{\scriptsize Full hypothesis space}} & \multirow{1}{*}{{\scriptsize $\frac{R_{p}}{R_{*}}$,}\textbf{\scriptsize{} }{\scriptsize $P_{\mathrm{clouds}}$,
$\mathbf{\boldsymbol{\xi}}=\mathrm{clr(}X_{\mathrm{H_{2}O}},X_{\mathrm{CO_{2}}},X_{\mathrm{CH_{4}}},X_{\mathrm{CO}},X_{\mathrm{H_{2}/He}},X_{\mathrm{N_{2}}})$}} & \multirow{1}{*}{{\scriptsize -284.57}} & \multirow{1}{*}{{\scriptsize 55.17}} & \multirow{1}{*}{{\scriptsize Ref.}}\tabularnewline
{\scriptsize $\mathrm{H_{2}O}$ removed} & {\scriptsize $\frac{R_{p}}{R_{*}}$,}\textbf{\scriptsize{} }{\scriptsize $P_{\mathrm{clouds}}$,
$\mathbf{\boldsymbol{\xi}}=\mathrm{clr(}X_{\mathrm{CO_{2}}},X_{\mathrm{CH_{4}}},X_{\mathrm{CO}},X_{\mathrm{H_{2}+He}},X_{\mathrm{N_{2}}})$} & {\scriptsize -292.21} & {\scriptsize 69.90} & \textbf{\scriptsize $B_{\mathrm{H_{2}O}}=2060\,\left(4.3\sigma\right)$}\tabularnewline
{\scriptsize $\mathrm{CO_{2}}$ removed} & {\scriptsize $\frac{R_{p}}{R_{*}}$,}\textbf{\scriptsize{} }{\scriptsize $P_{\mathrm{clouds}}$,
$\mathbf{\boldsymbol{\xi}}=\mathrm{clr(}X_{\mathrm{H_{2}O}},X_{\mathrm{CH_{4}}},X_{\mathrm{CO}},X_{\mathrm{H_{2}/He}},X_{\mathrm{N_{2}}})$} & {\scriptsize -284.63} & {\scriptsize 58.23} & {\scriptsize $B_{\mathrm{CO_{2}}}=$1.06}\tabularnewline
{\scriptsize $\mathrm{CH_{4}}$ removed} & {\scriptsize $\frac{R_{p}}{R_{*}}$,}\textbf{\scriptsize{} }{\scriptsize $P_{\mathrm{clouds}}$,
$\mathbf{\boldsymbol{\xi}}=\mathrm{clr(}X_{\mathrm{H_{2}O}},X_{\mathrm{CO_{2}}},X_{\mathrm{CO}},X_{\mathrm{H_{2}/He}},X_{\mathrm{N_{2}}})$} & {\scriptsize -284.31} & {\scriptsize 55.04} & {\scriptsize $B_{\mathrm{CH_{4}}}=$0.77}\tabularnewline
{\scriptsize CO removed} & {\scriptsize $\frac{R_{p}}{R_{*}}$, $P_{\mathrm{clouds}}$, $\mathbf{\boldsymbol{\xi}}=\mathrm{clr(}X_{\mathrm{H_{2}O}},X_{\mathrm{CO_{2}}},X_{\mathrm{CH_{4}}},X_{\mathrm{H_{2}/He}},X_{\mathrm{N_{2}}})$} & {\scriptsize -284.43} & {\scriptsize 55.31} & {\scriptsize $B_{\mathrm{CO}}=$0.86}\tabularnewline
{\scriptsize $\mathrm{N_{2}}$ removed} & {\scriptsize $\frac{R_{p}}{R_{*}}$,}\textbf{\scriptsize{} }{\scriptsize $P_{\mathrm{clouds}}$,
$\mathbf{\boldsymbol{\xi}}=\mathrm{clr(}X_{\mathrm{H_{2}O}},X_{\mathrm{CO_{2}}},X_{\mathrm{CH_{4}}},X_{\mathrm{CO}},X_{\mathrm{N_{2}}})$} & {\scriptsize -284.31} & {\scriptsize 56.61} & {\scriptsize $B_{\mathrm{N_{2}}}=$0.77}\tabularnewline
{\scriptsize $\mathrm{H_{2}/He}$ mix removed} & {\scriptsize $\frac{R_{p}}{R_{*}}$,}\textbf{\scriptsize{} }{\scriptsize $P_{\mathrm{clouds}}$,
$\mathbf{\boldsymbol{\xi}}=\mathrm{clr(}X_{\mathrm{H_{2}O}},X_{\mathrm{CO_{2}}},X_{\mathrm{CH_{4}}},X_{\mathrm{CO}},X_{\mathrm{H_{2}/He}})$} & {\scriptsize -284.50} & {\scriptsize 55.12} & {\scriptsize $B_{\mathrm{H_{2}/He}}=$0.93}\tabularnewline
{\scriptsize $\mathrm{H_{2}O}$ \& $\mathrm{CH_{4}}$ removed} & {\scriptsize $\frac{R_{p}}{R_{*}}$,}\textbf{\scriptsize{} }{\scriptsize $P_{\mathrm{clouds}}$,
$\mathbf{\boldsymbol{\xi}}=\mathrm{clr(}X_{\mathrm{CO_{2}}}X_{\mathrm{CH_{4}}},X_{\mathrm{CO}},X_{\mathrm{H_{2}/He}},X_{\mathrm{N_{2}}})$} & {\scriptsize -295.21} & {\scriptsize 83.49} & {\scriptsize $B_{\mathrm{H_{2}O\, or\, CH_{4}}}=$}\textbf{\scriptsize $41756\,\left(5.0\sigma\right)$}\tabularnewline
{\scriptsize Clouds removed} & {\scriptsize $\frac{R_{p}}{R_{*}}$, $\mathbf{\boldsymbol{\xi}}=\mathrm{clr(}X_{\mathrm{H_{2}O}},X_{\mathrm{CO_{2}}},X_{\mathrm{CH_{4}}},X_{\mathrm{CO}},X_{\mathrm{H_{2}/He}},X_{\mathrm{N_{2}}})$} & {\scriptsize -283.63} & {\scriptsize 55.34} & {\scriptsize $B_{\mathrm{Clouds}}=$0.39}\tabularnewline
\hline 
\end{tabular}
\par\end{centering}{\scriptsize \par}

\caption{Results of Bayesian model comparison for synthetic 80~ppm observations
of the water-dominated scenario on GJ~1214b. The synthetic observations
are depicted in Figure \ref{fig:Detection-of-water}(a). Values that
are directly referred to in Section 3 are marked in bold. The vector
$\mathbf{\boldsymbol{\xi}}$ is the center-log-ratio (clr) transformation
of the mole fractions of the atmospheric gases.\label{tab:Quantitative-model-comparison}}
\end{table*}

\subsubsection{Characteristic Degeneracy between the Mean Molecular Mass and Cloud
Top Pressure}

Transit observations with larger transit depth uncertainties, e.g.
80~ppm, can leave the water mole fraction unconstrained in the entire
range between 100\% and fractions of 1\% despite a robust $B_{\mathrm{H_{2}O}}=2040\,\left(4.3\sigma\right)$
detection of water absorption (Figure \ref{fig:Detection-of-water}).
High posterior probabilities would exist along a line of constant
feature depth, resulting in a degeneracy between the mean molecular
mass and the cloud top pressure. This degeneracy is characteristic
for transmission spectra that enable the detection of the absorber,
but do not provide sufficient constraints on the steepness of the
feature wings or depths of other features to constrain the mean molecular
mass independently.

In the hydrogen-rich regime, the feature size remains constant along
lines of constant water column density. The cloud top pressure compatible
with the synthetic observations decreases as the water mole fraction
in the atmosphere is increased (negative correlation). In the transition
region between 10\% and 90\% water mole fraction, the change in the
mean molecular mass dominates the change in the feature size. Maintaining
the same feature sizes requires higher cloud top pressures as the
water fraction is increased (positive correlation). Changes in the
feature sizes are small if the water mole fraction is increased because
the mean molecular mass remains largely unchanged and deeper layers
in the atmosphere are optically thick regardless of the exact water
mole fraction.

\subsubsection{Overlapping Molecular Bands in WFC3 Bandpass: $H_{2}O$ or $CH_{4}$
Absorption?}

A challenge in the spectral range of the \textit{HST~WFC3} bandpass
is that methane and water have two strongly overlapping absorption
bands at $1.15\,\mathrm{\mu m}$ and $1.4\,\mathrm{\mu m}$. An unambiguous
detection of water vapor requires not only the detection of the absorption
bands, but also requires one to determine whether the absorption was
caused by water, methane, or both. Distinction between the water and
methane cases is possible based on subtle differences in the absorption
feature shapes, the slope red of $1.6\,\mathrm{\mu m}$, and the different
opacities in the bandpass of the VLT observations. It is obvious,
however, that better observations are required to distinguish between
water features and similarly shaped methane features than is necessary
to distinguish between water features and a flat spectrum.

\begin{figure}[t]
\selectlanguage{british}%
\noindent \begin{raggedright}
\includegraphics[scale=0.32]{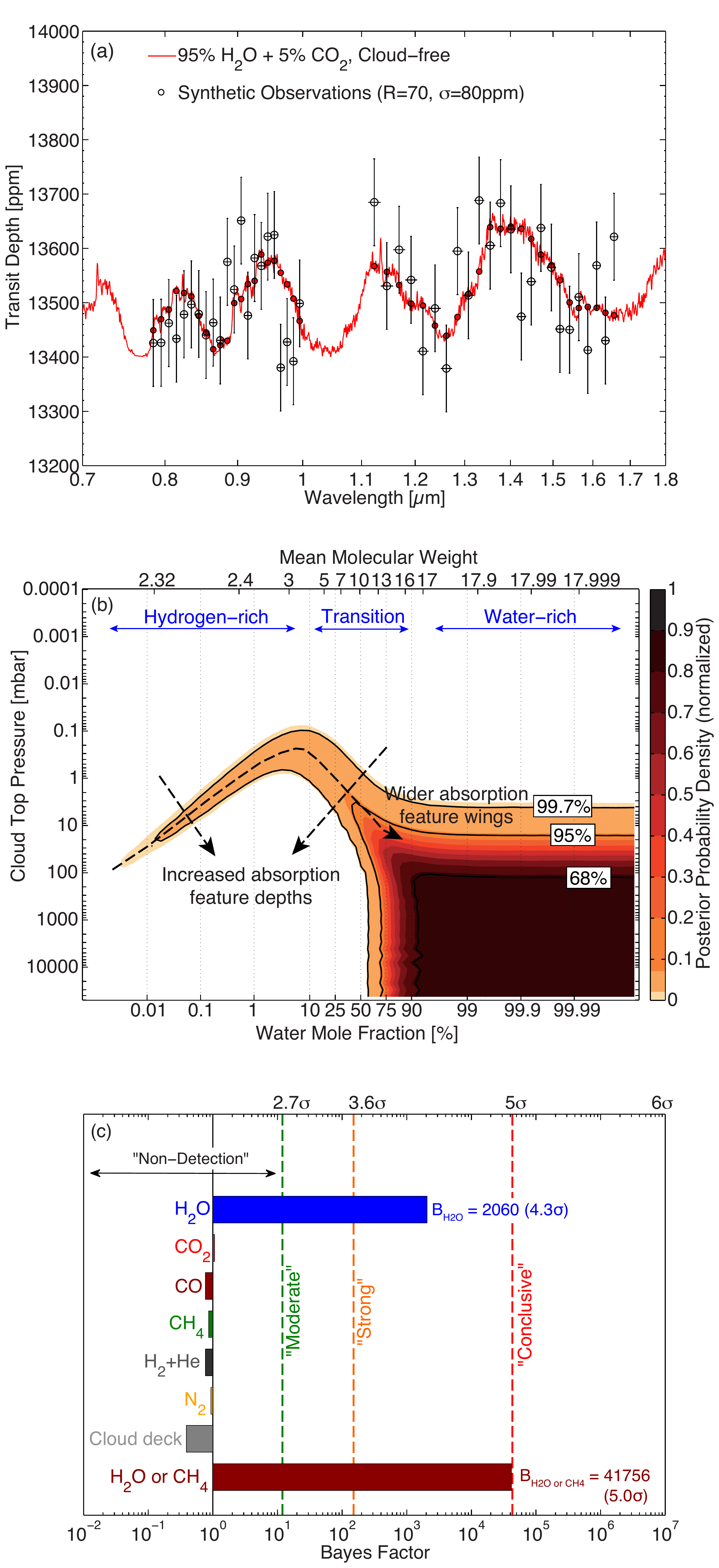}
\par\end{raggedright}

\selectlanguage{english}%
\caption{Synthetic 80~ppm observations and retrieval results for a cloud-free,
water-dominated atmosphere on GJ~1214b. The atmospheric scenario
and panels (a)-(c) are identical to the ones described Figure \ref{fig:H2O_Distinguish}.
80~ppm observations of a water-dominated atmosphere would lead to
a robust $ $detection of water vapor ($B_{\mathrm{H_{2}O}}=2060$),
but would be insufficient to robustly determine whether the atmosphere
is water-dominated or hydrogen-dominated. Atmospheric scenarios are
in agreement with the data along a contour of constant feature depth,
resulting in a degeneracy between the mean molecular mass and cloud
top pressure. The water mole fraction remains unconstrained between
0.02\% and 100\% at $3\sigma$. The degeneracy is characteristic for
transmission spectra that enable the detection of a single absorption,
but do not provide sufficient constraints on the steepness of the
feature wings or depth of other features to constrain the mean molecular
mass. \label{fig:Detection-of-water}}
\end{figure}

It is worth noting that the results from the Bayesian analysis in
this work inherently accounts for the overlap between water and methane
absorption features. The framework assigns a high probability for
the presence of water vapor only if the observed feature resembles
the expected shape for a water feature considerably closer than it
resembles the expected shape of a methane feature.  Taking again
the example of our 80~ppm synthetic observations (Figure \ref{fig:Detection-of-water}\textbf{),}
we find that our confidence in having detected water absorption is
lowered by the possibility that methane is responsible for the absorption
features. The Bayes factor as a measure of our confidence in the presence
of water is $B_{\mathrm{H_{2}O}}=2060\,\left(4.3\sigma\right)$ while
our confidence that either water or methane is present is $B_{\mathrm{H_{2}O\, or\, CH_{4}}}=41756\,\left(5.0\sigma\right)$.
In other words, it is easier to detect absorption features at $1.15\,\mathrm{\mu m}$
and $1.4\,\mathrm{\mu m}$ than it is to conclusively state that the
features were caused by water absorption, and this needs to be accounted
for when claiming the detection of water absorption. Another way to
look at it is that we could detect at $B=43000$ $\left(5\sigma\right)$
that either water or methane is present with 80~ppm observations,
but we need $\sim$70~ppm precision to conclude at the same confidence
that the feature was caused by water absorption.

\subsection{Effects of Non-Gray Aerosols\label{sub:Non-Gray-Aerosols-1}}

In Sections \ref{sub:Distinguishing-between-Water-Ric}-\ref{sub:Weak-Constraints-on},
we assumed gray clouds when demonstrating the distinction between
cloud-free water-rich atmospheres and cloudy hydrogen-rich atmospheres.
Here, we demonstrate that the fundamental approach of determining
the mean molecular mass based on wing slopes of molecular absorption
bands or by comparing features of the same absorber remains viable
in the general case of non-gray aerosols. We model cloud scattering
using Mie theory and find that the cores of molecular absorption bands
in transmission spectra are largely unaffected by the type and spectrum
of the particles. The molecular absorption bands therefore provide
unambiguous constraints on the mean molecular mass as long as there
are significant detectable molecular absorption features penetrating
the ``continuum'' spectrum of the haze or cloud particles.

\begin{figure*}[t]
\selectlanguage{british}%
\begin{centering}
\includegraphics[clip,width=0.7\textwidth]{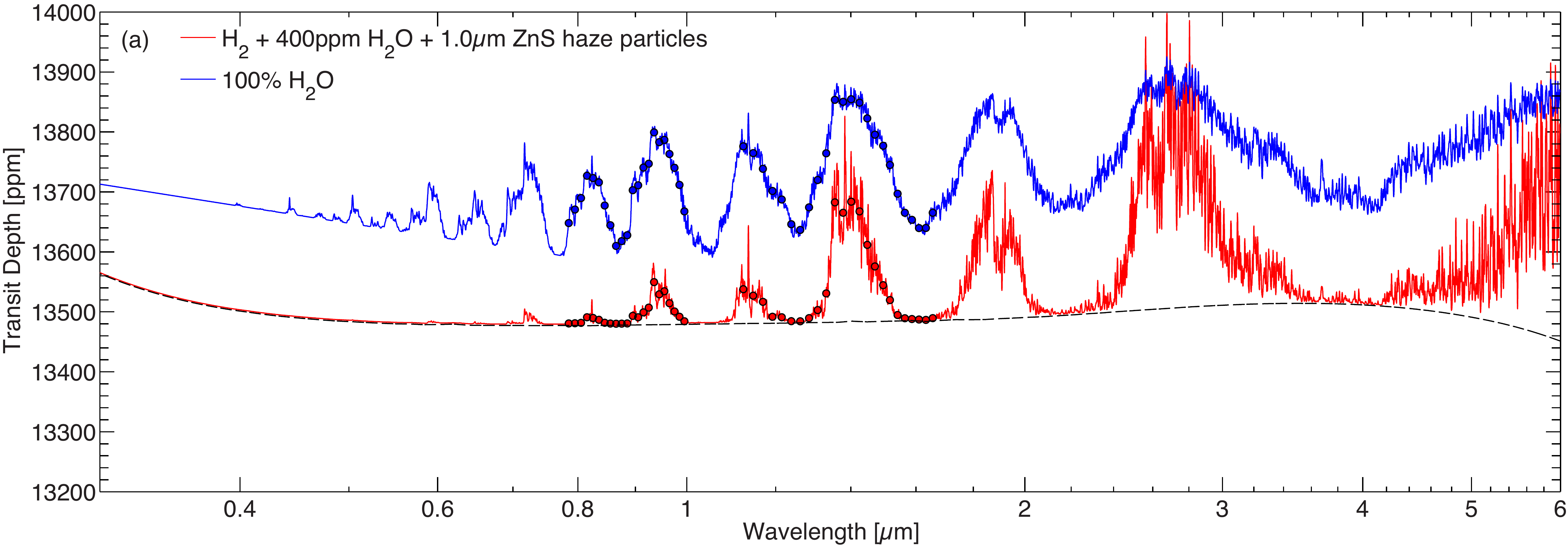}
\par\end{centering}

\begin{centering}
\includegraphics[clip,width=0.7\textwidth]{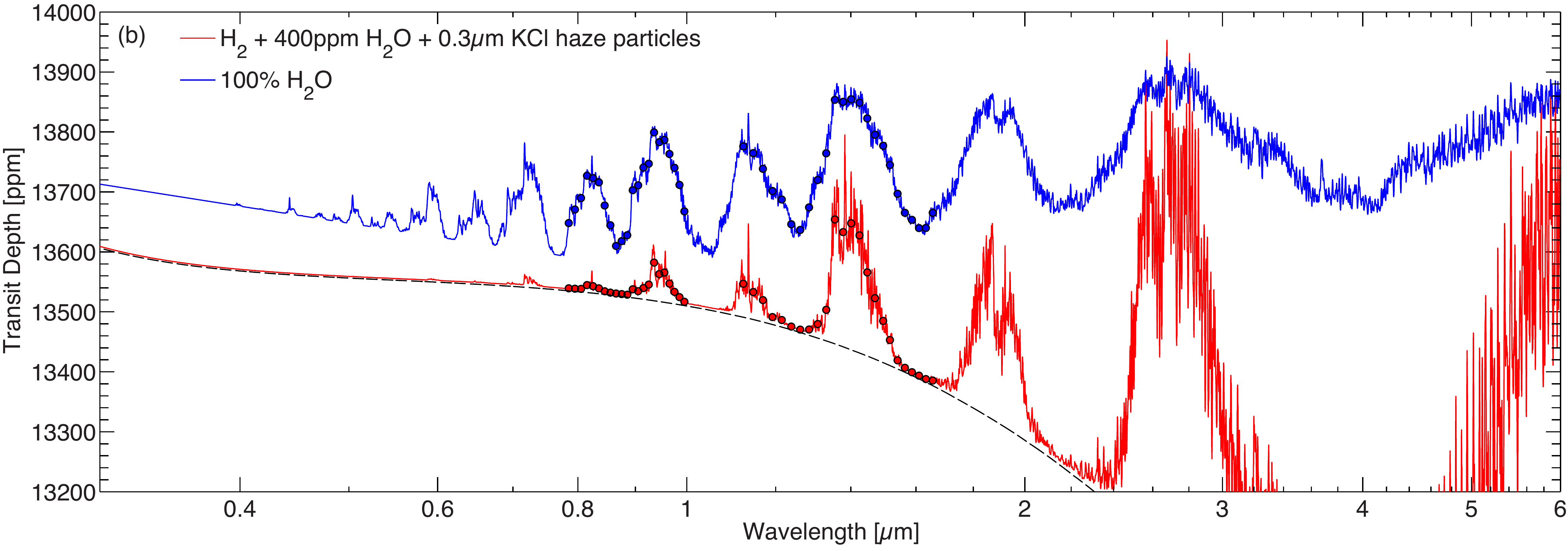}
\par\end{centering}

\selectlanguage{english}%
\caption[Comparison of model transmission spectra of water worlds and cloudy
sub-Neptunes with non-gray high-altitude clouds.]{Comparison of model transmission spectra of water worlds and cloudy
sub-Neptunes with non-gray high-altitude clouds. The red model spectra
in panels (a) and (b) are model spectra for two alternative scenarios
for hydrogen-rich atmospheres with thin, high-altitude hazes on GJ~1214b.
Panel (a) assumes ZnS particles with a mean radius of $\unit[1]{\text{\ensuremath{\mu}m}}$,
while panel (b) assumes KCl particles with a mean radius of $\unit[0.3]{\mu m}$.
The particles sizes and cloud top pressure (5~mbar) are chosen such
that the main water features in the HST~WFC3 are of similar size
as the ones predicted for water vapor atmospheres (blue spectra).
The non-gray effects of the hazes only are shown by the black dashed
spectra for which the gaseous absorption was set to zero. The cores
of the strong molecular absorption features remain largely unaffected
by the non-gray effects of the hazes. The steepness of the feature
wings and the relative depths between different absorption features
remain measures of mean molecular mass, even in the presence high-altitude
non-gray hazes, as long as significant absorption features can be
observed. The spectra are offset for clarity by slightly modifying
the planet's radius at the reference pressure level. \label{fig:ConstraintsHazes-1} }
\end{figure*}

We demonstrate that the wing slopes and relative sizes of water absorption
features are good measures of the mean molecular mass by presenting
two spectra of hydrogen-dominated atmospheres with high altitude hazes
and comparing them to cloud-free water atmospheres (Figure \ref{fig:ConstraintsHazes-1}).
We consider high-altitude haze particles composed of either condensed
Potassium Chloride (KCl) or Zinc Sulfide (ZnS) as two possible scenarios
that would sufficiently mute the spectral features of a hydrogen-dominated
atmosphere to match the published observations of GJ~1214b. KCl and
ZnS are considered for GJ~1214b because the he temperature pressure
profile in the atmosphere of GJ~1214b is likely to cross their condensation
curves, suggesting that KCl and ZnS may be able condense in the atmosphere
of GJ~1214b \citep{morley_neglected_2012,morley_quantitatively_2013}.

The particle size distributions and vertical extent of KCl and ZnS
condensates in the upper atmosphere of a super-Earths depend sensitively
on the atmospheric dynamics. Atmospheric dynamics on exoplanets are,
however, widely uncharacterized. The goal of this section is, therefore,
not to self-consistently model the size distributions and vertical
extent, but to demonstrate that the proposed approach to estimate
the mean molecular mass remains viable for non-gray clouds. 

For demonstrative purposes, we model thin, high-altitude KCl and ZnS
hazes using a standard analytical size distribution, a variation of
the gamma distribution introduced by \citep{hansen_multiple_1971},

\begin{equation}
n\left(r\right)=\mathrm{constant}\,\times\, r^{\left(1-3v_{\mathrm{eff}}\right)/v_{\mathrm{eff}}}e^{-r/r_{\mathrm{eff}}v_{\mathrm{eff}}}.
\end{equation}

The mean particle size $r_{\mathrm{eff}}$ is set to $\unit[0.3]{\mu m}$
and $\unit[1]{\mu m}$ for the KCl and ZnS haze scenarios, respectively.
The variance of the size distributions, $v_{\mathrm{eff}}$, is 0.1.
The hazes extend up to 5~mbar pressure level. The ratio of the condensed
mass and the gas mass in the upper atmosphere is 1 ppm. The vertical
extent of the hazes was chosen to obtain hydrogen-rich scenarios that
show water absorption features with transit depth variations similar
to those of water worlds, and, therefore, are most difficult to distinguish
from water-dominated scenarios. The particle sizes and condensed mass
fraction were selected to obtain haze spectra that considerably deviate
from the assumption of gray clouds while simultaneously providing
a reasonable fit to previously obtained transit depth measurements
of GJ~1214b. Particles smaller than $\unit[0.3]{\mu m}$ would increasingly
lead to steep slopes at near-infrared wavelengths. Larger particles
would lead to an increasingly gray appearance of the clouds in transmission,
and thus brings us back to the assumption made in Section \ref{fig:ConstraintsHazes-1}. 

Figure \ref{fig:ConstraintsHazes-1} shows that the cores of the strong
molecular absorption features are largely unaffected by the wavelength
dependence of the cloud/haze opacities. Estimates of the mean molecular
mass made based on the feature cores in transmission spectra are,
therefore, largely independent of the spectral properties of the cloud
properties. The cores of molecular absorption features in exoplanet
transmission spectra are largely independent of the cloud properties
for two reasons. First, for transmission spectra, the observed transit
depth at a given wavelength is almost exclusively determined by the
strongest opacity source at that wavelength. This is as a result of
the grazing geometry in which the transmission spectrum is formed
\citep{brown_transmission_2001}, combined with the exponential decrease
in gas density with altitude.  Second, aerosol opacities generally
change more gradually with wavelength than molecular opacities. Molecular
opacities at low pressures are dominated by sharp absorption lines
and bands that result from quantum mechanical transitions between
discrete vibrational and rotational states in the molecules. Light
extinction by aerosols, on the other hand, is a result of interference
of light that was scattered, refracted, or diffracted by a generally
continuous distribution of different particles sizes. Spectral features
arise as result of the interference or the wavelength dependence of
the complex refractive index, and the effects of the on the transmission
spectrum are more gradual than molecular state transitions. 

Second, while spectral features due to condensed phase absorption
can be prominent in reflective an thermal emission spectra, modeling
of exoplanet transmission spectra reveals that the features due to
condensed phase absorption are generally far less pronounced. The
reason for the difference is that multi-scattering of light can play
a dominant role in the approximately nadir-viewing geometry associated
with reflective and thermal emission spectra. The long pathways associated
with light that is scattered multiple times within in a cloud can
result in strong absorption features in reflective spectra of clouds.
Water clouds, for example, will appear highly reflective at visible
wavelengths for which the imaginary part of the refractive index is
low (low absorption), while it will appear almost black at some near-infrared
wavelength for which the imaginary part of the refractive index is
high (high absorption). In the exoplanet transit geometry, however,
any single scattering events will prevent the grazing light beams
from the host star to arrive at the observer due to large distance
between the target exoplanet and Earth.

\subsection{Probing the Composition of Volatile-Rich Atmospheres\label{sub:Probing-the-Molecular}}

The distinction between $\mathrm{H_{2}}$-dominated sub-Neptunes and
water or ice-rich worlds described in Section 3 is solely based on
the sharp contrast in molecular masses between hydrogen gas ($\mathrm{H_{2}}$)
and the ices ($\mathrm{H_{2}O}$, $\mathrm{CO_{2}}$, $\mathrm{CO}$,
$\mathrm{CH_{4}}$, $\mathrm{N_{2}}$, etc). The basic argumentation
is that primordial $\mathrm{H_{2}}$-dominated scenarios can be excluded
if the mean molecular mass deviates significantly from $\mu_{\mathrm{ave}}=2.3$.
Measuring a high mean molecular mass, however, does not unambiguously
determine the abundances of the individual volatile species in the
atmosphere. 

\begin{figure*}[t]
\selectlanguage{british}%
\noindent \begin{centering}
\includegraphics[clip,width=0.7\textwidth]{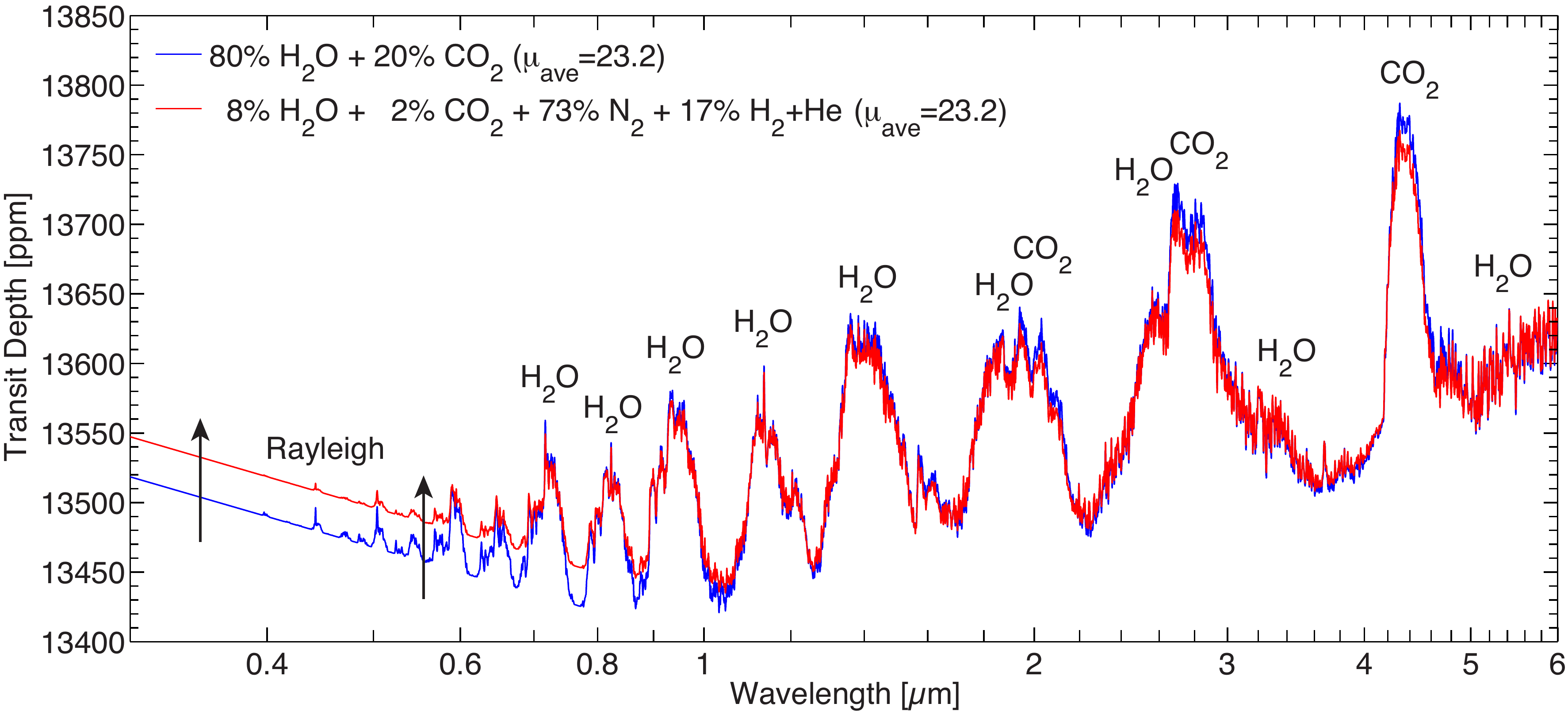}
\par\end{centering}

\selectlanguage{english}%
\caption{Challenge in distinguishing water-dominated atmospheres (80\%~$\mathrm{H_{2}O}$)
and water-rich atmospheres (80\%~$\mathrm{H_{2}O}$) with similar
mean molecular masses. Despite the vastly different compositions,
the $\mathrm{H_{2}O}$/$\mathrm{CO_{2}}$-dominated atmosphere (80\%
$\mathrm{H_{2}O}$ and 20\% $\mathrm{CO_{2}}$) and $\mathrm{H_{2}O}$/$\mathrm{CO_{2}}$-rich,
but $\mathrm{N_{2}}$-dominated atmosphere (8\% $\mathrm{H_{2}O}$
and 2\% $\mathrm{CO_{2}}$) show virtually identical infrared transmission
spectra. Atmospheric sceanrios with the same mean molecular masses
and the same relative abundances of the absorbers cannot be distinguished
using moderate-resolution NIR transmission spectra alone \citep{benneke_atmospheric_2012}.
Distinction is possible only at short wavelengths ($\lambda<\unit[1]{\mu m}$)
based on the increased Rayleigh scattering due to the presence of
$\mathrm{N_{2}}$ and $\mathrm{H_{2}+He}$. This is an example that
even a strong detection of a water feature is not sufficient to determine
whether the atmosphere is dominated by water vapor or only a small
fraction of the gas is water vapor.\label{fig:Relative-and-absolute} }
\end{figure*}

Figure \ref{fig:Relative-and-absolute} illustrates the difficulty
of determining the mole fraction of the individual ice species for
two vastly different atmospheres with identical molecular masses ($\mu_{ave}=23.2$).
Following the argumentation in \citet{benneke_atmospheric_2012},
the relative abundances of the ices that have IR absorption features,
e.g., $\mathrm{H_{2}O}$, $\mathrm{CO_{2}}$, $\mathrm{CO}$, $\mathrm{CH_{4}}$,
and $\mathrm{NH_{3}}$ can be determined by comparing the transit
depths in the strongest absorption bands of the different ices. An
atmosphere with 80\% $\mathrm{H_{2}O}$ + 20\% $\mathrm{CO_{2}}$
($\mathrm{H_{2}O}$/$\mathrm{CO_{2}}=4$), for example, can be distinguished
from an atmosphere with 90\% $\mathrm{H_{2}O}$ and 10\% $\mathrm{CO_{2}}$
($\mathrm{H_{2}O}$/$\mathrm{CO_{2}}=9$) because the transit depth
within the $\mathrm{CO_{2}}$ band at $\unit[4.3]{\mu m}$ would be
higher relative to the transit depths in the $\mathrm{H_{2}O}$ bands.

Scenarios with similar relative abundances of absorbing gases, however,
are practically indistinguishable through moderate-resolution ($R\sim100$)
infrared observations. Figure \ref{fig:Relative-and-absolute} shows
that the NIR spectrum of a volatile-dominated atmosphere can remain
virtually unchanged when the mole fraction of water is reduced from
80\% to 8\%. Spectrally inactive gases may be present in the correct
ratios for the mean molecular mass to remain unchanged. A distinction
between 80\% water and 8\% water therefore requires observations at
short wavelengths.

\section{Results: Analysis of Published HST WFC3 Spectra of GJ~1214b\label{sec:Results:-Analysis-of}}

\begin{figure}[h]
\includegraphics[clip,width=1\columnwidth]{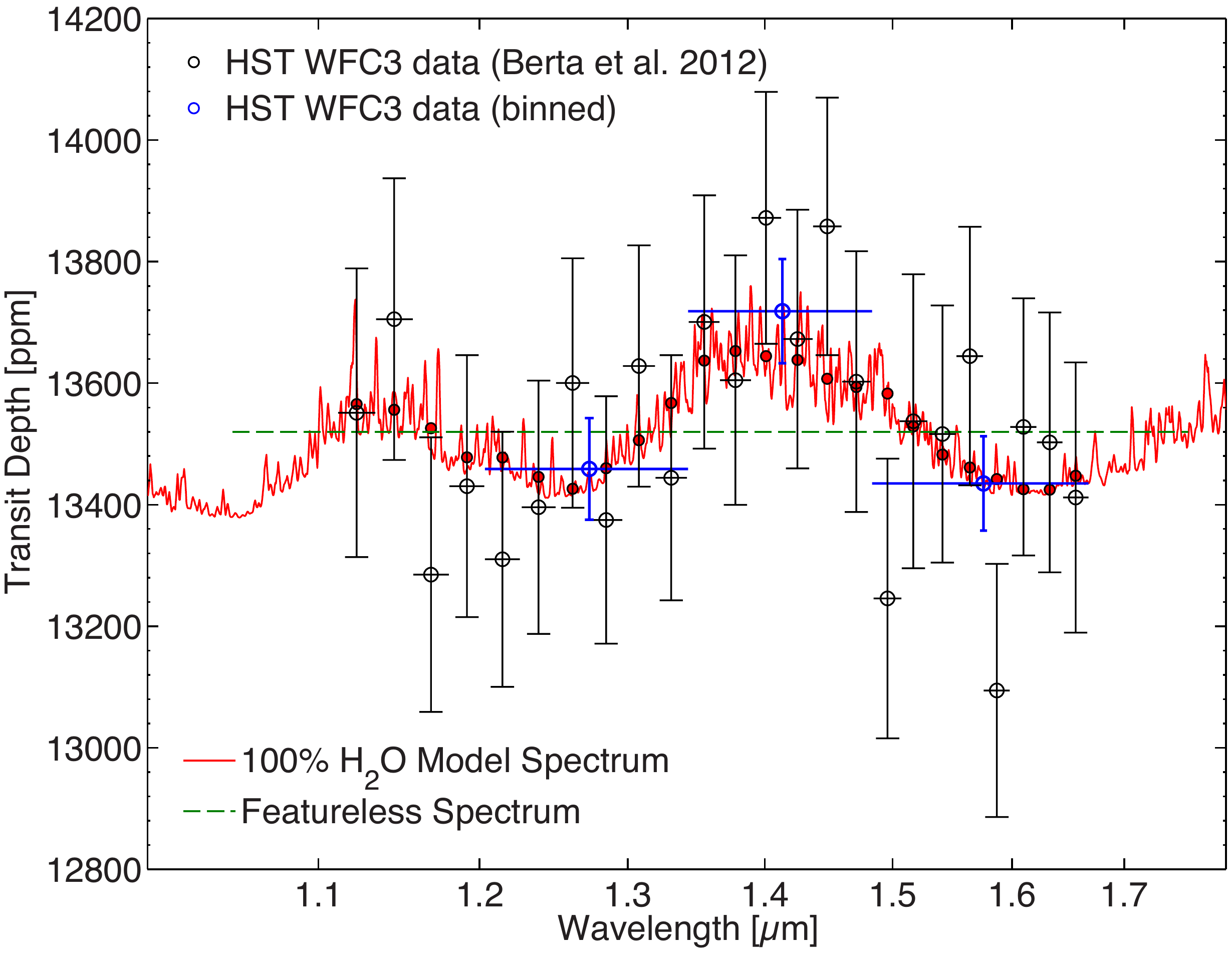}

\caption{Observed \textit{HST WFC3} observation of GJ~1214b by \citet{berta_flat_2012}
compared to model spectra. A featureless spectrum (green) and a model
spectrum of a 100\% water vapor atmosphere (red) are shown. Binned
data points (blue) are illustrated to indicate the proposed transit
depth variation near the $\unit[1.38]{\mu m}$ water feature. The
binned data points would suggest a $2.44\sigma$. The Bayesian analysis,
however, indicates that there is there is little statistical evidence
of water absorption in the \textit{HST WFC3} observations. We emphasize
that model-independent approaches to detect molecular absorption features
by binning data points ``inside'' the suspected feature and comparing
the transit depth to the surrounding ``continuum'' generally lead
to ambiguous results because spectra of thick atmospheres lack clear
separations between features and surrounding ``continuum'' (Table~\ref{tab:Molecular-abundances-of}).\label{fig:Spectrum} }
\end{figure}

A careful look at the observed \textit{HST WFC3} spectrum by \citet{berta_flat_2012}
reveals that the transit depth variation resembles the trends expected
for a clear, water-dominated atmosphere (Figure~\ref{fig:Spectrum}).
The resemblance was also pointed out by \citet{howe_theoretical_2012},
but no assessment of the significance of the water absorption was
provided.  In this Section, we assess the significance of the trend
in the transit depth variation in a statistically robust Bayesian
way. We also assess which constraints on cloud-top pressure and water/hydrogen
abundances can be derived from the $HST$ observation by \citet{berta_flat_2012}.
We demonstrate that the standard frequentist hypothesis testing based
on attempting to reject the ``flat spectrum'' null-hypothesis or
comparing data points ``within'' and ``outside'' the suggested
features generally leads to ambiguous or misleading results.

\subsection{Testing for the Presence of Water Absorption\label{sub:Testing-the-Presence}}

Our Bayesian analysis shows that, despite the suggested trend in the
spectral data points, there is little statistical evidence of water
absorption in the \textit{HST WFC3} observations by \citet{berta_flat_2012}
(Table~\ref{tab:ModelComp_Berta2012}). The Bayesian factor $B_{\mathrm{H_{2}O}}$
describing our confidence in the presence of water is only 3.34, which
is at best a weak suggestion that water absorption may be present.
The Bayes factor for all other molecule and clouds are below $\sim2$,
indicating that no molecular absorption can be inferred from the data
set.

It is important to note that standard model-independent approaches
based on binning data points ``within'' and ``outside'' can lead
to ambiguous or even misleading results for a spectral data set such
as that provided by \citet{berta_flat_2012}. The reason is that molecular
absorption features of thick atmospheres are generally not box-shaped.
The gradual shape of the absorption feature wings make it impossible
to unambiguously assign which data points are ``inside'' and ``outside''
of the water absorption features. Figure~\ref{fig:Spectrum} and
Table \ref{tab:Molecular-abundances-of} show that the contrast between
binned data points ``inside'' and ``outside'' can vary between
$\sim1.4\sigma$ and $\sim\mathrm{2.65\sigma}$, depending on which
data points are considered within the feature. A finding of an absorption
feature at $2.65\sigma$ could then be misinterpreted as strong suggestion
of water absorption. 

The Bayesian framework presented in this work does not lead to ambiguity
because the data are taken as given. High confidence in the detection
of water absorption is only assigned if the original data points follow
the spectral shapes expected for atmospheres with water absorption,
and no other molecular absorber can explain the data.

\begin{table*}[t]
\noindent \begin{centering}
{\scriptsize }%
\begin{tabular}{llccc}
\hline 
{\scriptsize Retrieval model} & {\scriptsize Retrieval model parameters} & {\scriptsize Evidence } & {\scriptsize Best-fit} & {\scriptsize Bayes factor }\tabularnewline
 &  & {\scriptsize $\ln\left(\mathcal{Z}_{i}\right)$} & {\scriptsize $\chi_{\mathrm{best-fit}}^{2}$} & {\scriptsize $B_{i}=\mathcal{Z}_{0}/\mathcal{Z}_{i}$}\tabularnewline
\hline 
\multirow{1}{*}{{\scriptsize Full hypothesis space}} & \multirow{1}{*}{{\scriptsize $\frac{R_{p}}{R_{*}}$,}\textbf{\scriptsize{} }{\scriptsize $P_{\mathrm{clouds}}$,
$\mathbf{\boldsymbol{\xi}}=\mathrm{clr(}X_{\mathrm{H_{2}O}},X_{\mathrm{CO_{2}}},X_{\mathrm{CH_{4}}},X_{\mathrm{CO}},X_{\mathrm{H_{2}/He}},X_{\mathrm{N_{2}}})$}} & \multirow{1}{*}{{\scriptsize -162.96}} & \multirow{1}{*}{{\scriptsize 12.24}} & \multirow{1}{*}{{\scriptsize Ref.}}\tabularnewline
{\scriptsize $\mathrm{H_{2}O}$ removed} & {\scriptsize $\frac{R_{p}}{R_{*}}$,}\textbf{\scriptsize{} }{\scriptsize $P_{\mathrm{clouds}}$,
$\mathbf{\boldsymbol{\xi}}=\mathrm{clr(}X_{\mathrm{CO_{2}}},X_{\mathrm{CH_{4}}},X_{\mathrm{CO}},X_{\mathrm{H_{2}+He}},X_{\mathrm{N_{2}}})$} & {\scriptsize -163.16} & {\scriptsize 14.31} & \textbf{\scriptsize $B_{\mathrm{H_{2}O}}=3.34$}\tabularnewline
{\scriptsize $\mathrm{CO_{2}}$ removed} & {\scriptsize $\frac{R_{p}}{R_{*}}$,}\textbf{\scriptsize{} }{\scriptsize $P_{\mathrm{clouds}}$,
$\mathbf{\boldsymbol{\xi}}=\mathrm{clr(}X_{\mathrm{H_{2}O}},X_{\mathrm{CH_{4}}},X_{\mathrm{CO}},X_{\mathrm{H_{2}/He}},X_{\mathrm{N_{2}}})$} & {\scriptsize -164.16} & {\scriptsize 12.20} & {\scriptsize $B_{\mathrm{CO_{2}}}=$1.26}\tabularnewline
{\scriptsize $\mathrm{CH_{4}}$ removed} & {\scriptsize $\frac{R_{p}}{R_{*}}$,}\textbf{\scriptsize{} }{\scriptsize $P_{\mathrm{clouds}}$,
$\mathbf{\boldsymbol{\xi}}=\mathrm{clr(}X_{\mathrm{H_{2}O}},X_{\mathrm{CO_{2}}},X_{\mathrm{CO}},X_{\mathrm{H_{2}/He}},X_{\mathrm{N_{2}}})$} & {\scriptsize -163.19} & {\scriptsize 12.18} & {\scriptsize $B_{\mathrm{CH_{4}}}=$1.63}\tabularnewline
{\scriptsize CO removed} & {\scriptsize $\frac{R_{p}}{R_{*}}$, $P_{\mathrm{clouds}}$, $\mathbf{\boldsymbol{\xi}}=\mathrm{clr(}X_{\mathrm{H_{2}O}},X_{\mathrm{CO_{2}}},X_{\mathrm{CH_{4}}},X_{\mathrm{H_{2}/He}},X_{\mathrm{N_{2}}})$} & {\scriptsize -163.44} & {\scriptsize 12.28} & {\scriptsize $B_{\mathrm{CO}}=$1.70}\tabularnewline
{\scriptsize $\mathrm{N_{2}}$ removed} & {\scriptsize $\frac{R_{p}}{R_{*}}$,}\textbf{\scriptsize{} }{\scriptsize $P_{\mathrm{clouds}}$,
$\mathbf{\boldsymbol{\xi}}=\mathrm{clr(}X_{\mathrm{H_{2}O}},X_{\mathrm{CO_{2}}},X_{\mathrm{CH_{4}}},X_{\mathrm{CO}},X_{\mathrm{N_{2}}})$} & {\scriptsize -163.49} & {\scriptsize 12.72} & {\scriptsize $B_{\mathrm{N_{2}}}=$1.82}\tabularnewline
{\scriptsize $\mathrm{H_{2}/He}$ mix removed} & {\scriptsize $\frac{R_{p}}{R_{*}}$,}\textbf{\scriptsize{} }{\scriptsize $P_{\mathrm{clouds}}$,
$\mathbf{\boldsymbol{\xi}}=\mathrm{clr(}X_{\mathrm{H_{2}O}},X_{\mathrm{CO_{2}}},X_{\mathrm{CH_{4}}},X_{\mathrm{CO}},X_{\mathrm{H_{2}/He}})$} & {\scriptsize -163.56} & {\scriptsize 12.45} & {\scriptsize $B_{\mathrm{H_{2}/He}}=$2.01}\tabularnewline
{\scriptsize $\mathrm{H_{2}O}$ \& $\mathrm{CH_{4}}$ removed} & {\scriptsize $\frac{R_{p}}{R_{*}}$,}\textbf{\scriptsize{} }{\scriptsize $P_{\mathrm{clouds}}$,
$\mathbf{\boldsymbol{\xi}}=\mathrm{clr(}X_{\mathrm{CO_{2}}}X_{\mathrm{CH_{4}}},X_{\mathrm{CO}},X_{\mathrm{H_{2}/He}},X_{\mathrm{N_{2}}})$} & {\scriptsize -163.65} & {\scriptsize 17.56} & {\scriptsize $B_{\mathrm{H_{2}O\, or\, CH_{4}}}=3.65$}\tabularnewline
{\scriptsize Clouds removed} & {\scriptsize $\frac{R_{p}}{R_{*}}$, $\mathbf{\boldsymbol{\xi}}=\mathrm{clr(}X_{\mathrm{H_{2}O}},X_{\mathrm{CO_{2}}},X_{\mathrm{CH_{4}}},X_{\mathrm{CO}},X_{\mathrm{H_{2}/He}},X_{\mathrm{N_{2}}})$} & {\scriptsize -164.25} & {\scriptsize 12.51} & {\scriptsize $B_{\mathrm{Clouds}}=$1.23}\tabularnewline
\hline 
\end{tabular}
\par\end{centering}{\scriptsize \par}

\caption{Results of Bayesian model comparison for observed \textit{HST WFC3}
transmission spectrum of GJ~1214b by \citet{berta_flat_2012}. The
observations are depicted in Figure~\ref{fig:Spectrum}(a).\label{tab:ModelComp_Berta2012}}
\end{table*}

\begin{table*}[t]
\begin{centering}
{\scriptsize }%
\begin{tabular}{l>{\centering}p{4cm}cccccc}
\hline 
\noalign{\vskip\doublerulesep}
{\scriptsize Binning scheme} & {\scriptsize Indices of data points}\linebreak {\scriptsize \textquotedbl{}inside\textquotedbl{}
the feature} & \multicolumn{2}{>{\centering}p{4cm}}{{\scriptsize Indices of data points}\linebreak {\scriptsize in surrounding
\textquotedbl{}continuum\textquotedbl{}}} & \multicolumn{2}{c}{{\scriptsize 1 - p-Value}} & \multicolumn{2}{>{\centering}p{4cm}}{{\scriptsize Detection}\linebreak {\scriptsize significance}}\tabularnewline[\doublerulesep]
\hline 
\noalign{\vskip\doublerulesep}
\noalign{\vskip\doublerulesep}
{\scriptsize \#1 (Figure \ref{fig:Spectrum})} & {\scriptsize 11-16} & {\scriptsize 5-10 (left)} & {\scriptsize 17-24 (right)} & {\scriptsize 96.9\%} & {\scriptsize 98.5\%} & {\scriptsize $2.16\,\sigma$ (left)} & {\scriptsize $2.44\,\sigma$ (right)}\tabularnewline[\doublerulesep]
\noalign{\vskip\doublerulesep}
\noalign{\vskip\doublerulesep}
{\scriptsize \#2} & {\scriptsize 10-17} & {\scriptsize 5-9 (left)} & {\scriptsize 18-24 (right)} & {\scriptsize 82.9\%} & {\scriptsize 85.6\%} & {\scriptsize $1.37\,\sigma$ (left)} & {\scriptsize $1.46\,\sigma$ (right)}\tabularnewline[\doublerulesep]
\noalign{\vskip\doublerulesep}
\noalign{\vskip\doublerulesep}
{\scriptsize \#3} & {\scriptsize 11-16} & \multicolumn{2}{c}{{\scriptsize 5-10 and 17-24}} & \multicolumn{2}{c}{{\scriptsize 99.2\%}} & \multicolumn{2}{c}{{\scriptsize $2.65\,\sigma$}}\tabularnewline[\doublerulesep]
\noalign{\vskip\doublerulesep}
\noalign{\vskip\doublerulesep}
{\scriptsize \#4} & {\scriptsize 10-17} & \multicolumn{2}{c}{{\scriptsize 5-9 and 18-24}} & \multicolumn{2}{c}{{\scriptsize 90.5\%}} & \multicolumn{2}{c}{{\scriptsize $1.67\,\sigma$}}\tabularnewline[\doublerulesep]
\hline 
\noalign{\vskip\doublerulesep}
\end{tabular}
\par\end{centering}{\scriptsize \par}

\caption{Ambiguity in the model-independent detection significances for non-box-shaped
spectral features. The table presents the detection significances
of an increased transit depth in the $\unit[1.38]{\mu m}$ water band
derived from the published transmission spectrum of GJ~1214b by \citet{berta_flat_2012}.
The indices of the data points are counted from the shortest wavelength
data point \#1 at $\unit[1.123]{\mu m}$ to the longest data point
\#24 at $\unit[1.656]{\mu m}$ (Figure~\ref{fig:Spectrum}). For
thick atmospheres, there is no clear cutoff to determine which data
points belong to the feature and which do not. The model-independent
detection significance ranges between 1.46$\sigma$ and $2.65\sigma$
depending on which data points are considered to be inside the water
band and which data points are considered to be in the surrounding
continuum. \label{tab:Molecular-abundances-of}}
\end{table*}

\subsection{Atmospheric Constraints}

\begin{figure*}[t]
\noindent \begin{centering}
\includegraphics[clip,width=0.7\textwidth]{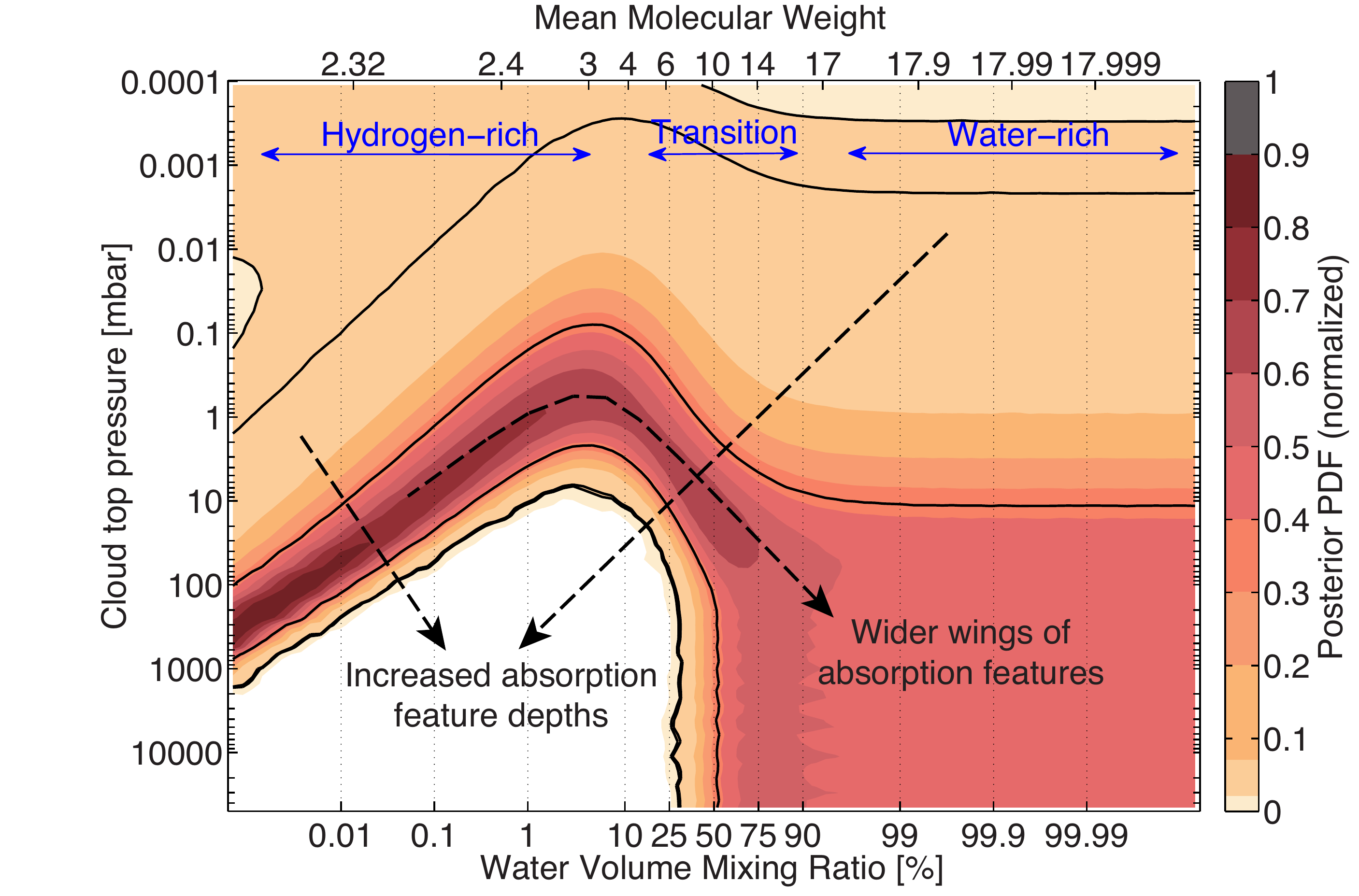}
\par\end{centering}

\caption{Atmospheric constraints derived from the observed transmission spectrum
by \citet{berta_flat_2012}. The shading illustrates the marginalized
posterior probability as a function of mean molecular mass and cloud
top pressure for a two-component $\mathrm{H_{2}O}+\mathrm{H_{2}}$
atmosphere. The black contour lines indicate the 68\%, 95\%, and 99.7\%
Bayesian credible regions. The 95\% Bayesian credible region extends
over large parts of the parameter space, preventing an unambiguous
characterization of the atmosphere of GJ~1214b with the currently
available data. Starting from the most robust statement, we can rule
out clear hydrogen-dominated atmospheres with hydrogen mole fractions
above 40\% at $>3\sigma$. Water-dominated and cloudy, hydrogen-dominated
atmospheres are, however, in agreement with the data at $1\sigma$
due to degeneracy between the water mole fraction and the cloud top
pressure (compare Figure \ref{fig:Detection-of-water}). The prior
probability is uniform in the parameter space spanned by $\log\left(p_{surf}\right)$,
$\xi_{1}=\log\left(X_{H2O}/\sqrt{X_{H2O}\cdot X_{H2}}\right)$.\label{fig:Marginalized-posterior-probabili-2}}
\end{figure*}

The observed transit spectrum by \citet{berta_flat_2012} leads to
correlated constraints on the mean molecular mass (or water mole fraction)
and the cloud top pressure that are characteristic for transit observations
that show a weak trend of an absorption feature (Figure \ref{fig:Marginalized-posterior-probabili-2}).
The highest posterior probabilities exist along a curve of constant
feature size corresponding to the apparent variation in the \citet{berta_flat_2012}
data. The 99.7\% ($3\sigma$) extends, however, all the way to virtually
featureless scenarios with low cloud top pressures below 0.0001~mbar. 

Cloud-free hydrogen-rich scenarios (bottom left region of Figure \ref{fig:Marginalized-posterior-probabili-2})
are disfavored by the observational data because the observed transmission
spectra do not show the deep absorption features expected for clear
hydrogen-rich atmospheres. At 99.7\% probability ($3\sigma$), we
conclude that cloud free atmospheres are only in agreement with the
data if the mean molecular mass is above 7, corresponding to more
than 30\% $\mathrm{H_{2}O}$ in an $\mathrm{H_{2}O}$-$\mathrm{H_{2}}$
atmosphere.

Hydrogen-dominated scenarios with less than 2\% water are possible
only if clouds are present at low pressure such that the partial pressure
of water vapor at the cloud surface does not exceed $p_{\mathrm{H_{2}O}}\sim0.025\,\mathrm{mbar}$.
Above 2\% water vapor, the water vapor significantly increases the
mean molecular mass, thereby reducing the size of the molecular absorption
features. As a result, the $3\sigma$ contour of the Bayesian credible
region peaks at $X_{\mathrm{H_{2}O}}\approx2\%$ and $p_{\mathrm{surf}}=10\,\mathrm{mbar}$
and then continues to fall to higher cloud top pressures as the water
mole fraction is increased further.

Our findings are in qualitative agreement with the list of implications
for the atmospheric composition provided by \citet{berta_flat_2012}.
One major difference is that we provide Bayesian credible regions
describing our knowledge about the atmospheric composition in a statistically
robust probabilistic way. \citet{berta_flat_2012} derived implications
on the atmospheric composition using $\chi^{2}$ hypothesis testing
and sequentially rejecting (or failing to reject) predefined atmospheric
scenarios.

In additional difference is that we used the new mass, radius, and
surface gravity estimates by Carter et al. (2013, in preparation).
The new estimate for the planet's surface gravity is 53\% higher than
the one published by \citet{charbonneau_super-earth_2009}. As a result,
the scale height and modeled feature depth for a given atmospheric
scenario decreases by 34.5\%.

\section{Discussion\label{sec:Discussion}}

\subsection{Scaling Laws for Required Observation Precision \label{sub:Scaling-Law-for}}

\paragraph{Planet and Star Scenario}

The quantitative results obtained for the super-Earth GJ~1214b in
Section~3 can be generalized for transiting super-Earth exoplanets
with different bulk properties and different host stars using the
scaling law

\begin{equation}
\frac{\sigma_{\mathrm{req}}}{\sigma_{\mathrm{ref}}}\thickapprox\frac{R_{P}}{R_{P,\text{ref}}}\frac{T_{eq}}{T_{eq,\text{ref}}}\frac{g_{\mathrm{p,ref}}}{g_{\mathrm{p}}}\frac{R_{\mathrm{*,ref}}^{2}}{R_{*}^{2}}.\label{eq:sigma_req}
\end{equation}

The required precision $\sigma_{\mathrm{req}}$ is the maximum allowed
observational uncertainty in the spectral transit depth measurements,
$R_{p}$ is the planetary radius, $T_{eq}$ is the equilibrium temperature,
$g_{\mathrm{p}}$ is the surface gravity of the planet, and $R_{*}$
is the radius of the host star. Equation (\ref{eq:sigma_req}) is
derived by relating the area $2R_{p}H$ of an annulus around the planet
and a width of one scale height $H\propto T_{eq}/g$ to the cross
sectional area of the stellar disk $\pi R_{*}^{2}$. 

Equation (\ref{eq:sigma_req}) is valid for clear, hazy, and cloudy
atmospheres to within a few percent error as long as identical scenarios
for the atmospheric compositions and clouds/hazes properties are compared.
This simple scaling law is valid for all types of atmospheres because
differences between the transit depths observed at two different wavelengths
are directly related to the difference in the altitudes $\Delta z$
at which the atmosphere becomes opaque to a grazing light beam at
the two wavelengths. The difference in the altitudes $\Delta z$,
in turn, is related to the pressure scale on which the atmospheric
scenarios, e.g., cloud top pressure and compositional profiles, are
defined by $\Delta z=-H\cdot\Delta P/P$. Equation~(\ref{eq:sigma_req})
does not account for the increase in atmospheric path length of grazing
light beams due to an increase in the planet radius. The longer path
length, however, generally leads to an approximately uniform increase
in transit depths across the entire spectrum and has little effect
on the transit depth variations.

Equation (\ref{eq:sigma_req}) shows that the required observational
precision depends most strongly on the radius of the host star due
to large differences in the stellar radii and the square dependence
on that parameter. The stellar radius of the M-dwarf GJ~1214 is only
18.8\% of the radius of a Sun-like star. As a result, the required
photometric precision to characterize a planet orbiting a Sun-like
star is $\sim$28 times larger than the required precision for the
characterization of the same planet orbiting GJ~1214. Atmospheric
characterizations of a planet orbiting a Sun-like star with the same
apparent brightness would therefore require more than 500 times more
observational time than a similar atmospheric characterization requires
for a planet orbiting GJ~1214. The effect of the atmospheric temperature
and planetary radius on the required precision scales only linearly.
The required precision for the atmospheric characterization of an
temperate super-Earth orbiting a nearby M-dwarf, for example, is only
higher by a factor of $2-3$ higher compared to GJ~1214b (see Section
\ref{sub:Atmospheric-Characterization-of} for a discussion). Similarly,
the required precision would only increase by a factor of 2.4 if an
Earth-sized planet was studied rather than GJ~1214b.

\paragraph{Spectral Resolution of Observations}

The quantitative results of the photometric precision in Section 3
are presented for a resolving power $R=\lambda/\Delta\lambda=70$.
Higher resolution spectra, whose uncertainties are dominated by white
noise, can provide the same information, even if the uncertainties
of the individual data points are higher. Assuming that the observational
uncertainties are dominated by white noise, the requirements on the
precision of each single data point scale as 

\begin{equation}
\sigma_{\mathrm{HR,\, req}}\thickapprox\sigma_{\mathrm{ref}}\sqrt{\frac{\Delta\lambda_{\mathrm{ref}}}{\Delta\lambda_{\mathrm{HR}}}}=\sigma_{\mathrm{ref}}\sqrt{\frac{R_{\mathrm{HR}}}{R_{\mathrm{ref}}}},\label{eq:sigma_req_resolving}
\end{equation}

for $R_{\mathrm{HR}}>R_{\mathrm{ref}}$. Equation \ref{eq:sigma_req_resolving}
describes in a formal way that the required precision, $\sigma_{\mathrm{HR,\, req}}$,
of spectra with a higher resolving power, $R_{\mathrm{HR}}$, can
be binned to reduce the observational uncertainty of individual data
points to $\sigma_{\mathrm{ref}}$ at a reference resolving power
of $R_{\mathrm{ref}}$. The higher resolution spectra, therefore,
contain at least as much information about the atmosphere as the spectra
with observational uncertainty $\sigma_{\mathrm{ref}}$ and reference
resolving power $R_{\mathrm{ref}}$.

Equation (\ref{eq:sigma_req_resolving}) can be regarded as a conservative
scaling law because observations with significantly higher resolving
power than $R=\lambda/\Delta\lambda=70$ may capture individual peaks
within the water absorption features that are not captured at low
resolution. High resolution observations may therefore provide additional
information to better constrain the gradient $dR_{P,\lambda}/d\text{\ensuremath{\left(\ln\sigma_{\lambda}\right)}}$
and thus the mean molecular mass (equation\ref{eq:1}). We confirm
in numerical studies that equation (\ref{eq:sigma_req_resolving})
is valid for resolving power between $\sim50$ and several hundreds.
A detailed numerical exploration of the effect of the resolving power
is beyond the scope of this study.

\subsection{HD~97658b, 55~Cnc~e, and GJ~436b\label{sub:HD97658b,-55Cnce,-and}}

Besides GJ~1214b, the transiting exoplanets in the mass regime of
super-Earths and Neptunes that are most amenable to study are HD~97658b,
55~Cnc~e, and GJ~436b. Their bulk densities are high enough to
require a larger ice or rock fraction than the solar system ice giants,
but far too low to be explained by an entirely Earth-like rocky composition.
Similar to GJ~1214b, the question arises whether these planets are
surrounded by a thick hydrogen-dominated envelope or a ice-dominated
gas envelope.

We find from Equation (\ref{eq:sigma_req}) that the precision required
to distinguish between water and hydrogen-rich scenarios for HD~97658b,
55~Cnc~e, and GJ~436b is significantly higher (Table ). The main
driver in increasing the required precision is the larger host star
diameters. It should be noted, however, the photon flux received from
HD~97658 and 55~Cnc is orders of magnitude higher due to the brightness
of the stars (Table \ref{tab:ScalingForHD97658b}). Significantly
higher transit depth precision may therefore be achievable per transit
if photon limited observations and a high integration efficiency can
be achieved. The implications of the host star brightness on the instrumental
effects and integration are specific to each instrument and beyond
the discussion in this work.

\begin{figure*}[t]
\begin{centering}
\includegraphics[clip,width=0.7\textwidth]{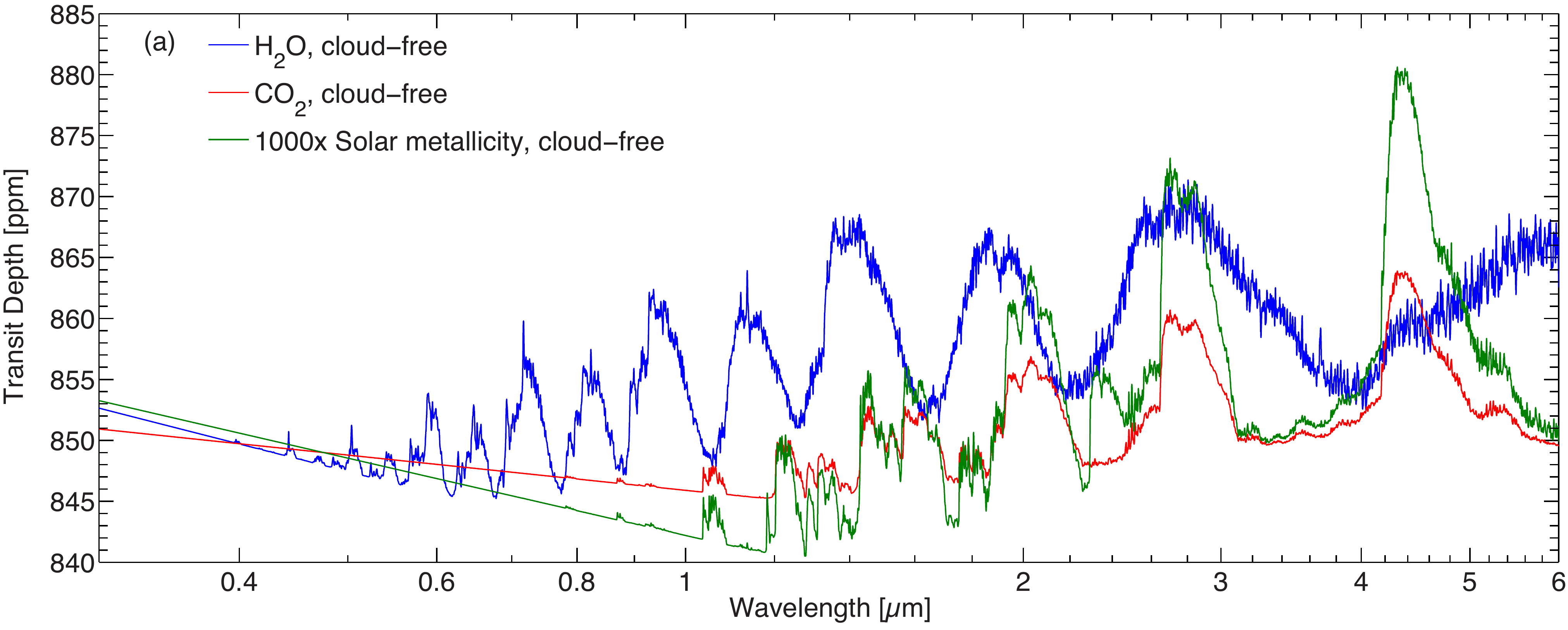}
\par\end{centering}

\begin{centering}
\includegraphics[clip,width=0.7\textwidth]{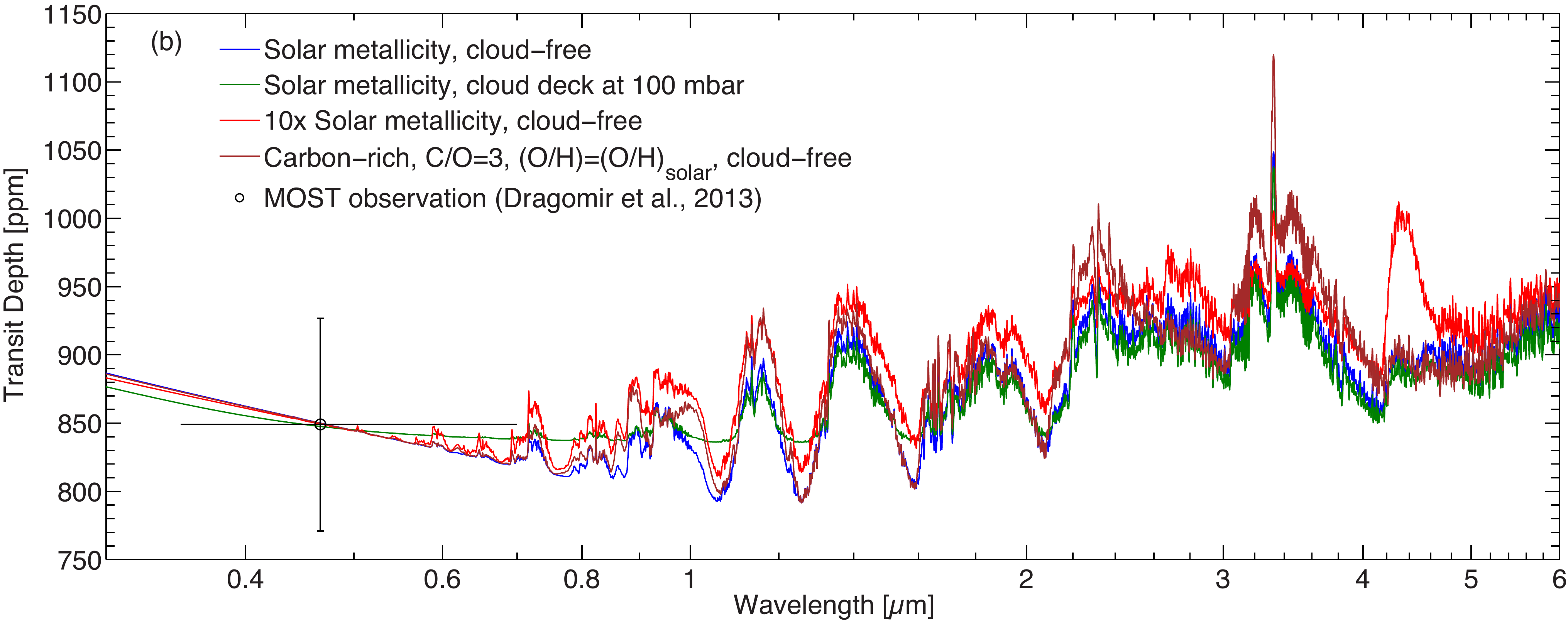}
\par\end{centering}

\caption{Model transmission spectra for the super-Earth HD~97658b. Panel (a)
shows spectra for atmospheres dominated by volatiles with high mean
molecular masses. Panel (b) shows spectra for hydrogen-dominated scenarios.
All spectra are modeled to match the transit depth measurement in
the \textit{MOST} bandpass by \citet{dragomir_new_2013} (black).
\label{fig:SpectraHD97658b}}
\end{figure*}

Transmission spectra of the exoplanet HD~97658b are displayed in
Figure \ref{fig:SpectraHD97658b} to further guide future observations.
The planet was recently found to transit are displayed \citep{henry_detection_2011,dragomir_new_2013}.
Cloud-free hydrogen-dominated atmospheres would results in transit
variation of $\unit[100-200]{ppm}$ that would readily be detectable
with currently available instrumentation. A confident detection of
water vapor or other volatile-rich atmospheric scenarios would require
a precision below $\unit[10]{ppm}$ at moderate spectral resolution
(Table \ref{tab:ScalingForHD97658b}). 

\begin{table*}[t]
\begin{centering}
\begin{tabular}[b]{cc>{\centering}m{4.5cm}>{\centering}m{4.5cm}c>{\centering}m{2cm}}
\hline 
\noalign{\vskip\doublerulesep}
\multicolumn{1}{l}{} & \multicolumn{1}{c}{} & \multicolumn{2}{c}{{\scriptsize Required precision $\sigma_{\mathrm{req}}$ at $R=70$}} & \multicolumn{2}{>{\centering}p{3cm}}{{\scriptsize Host star}\linebreak {\scriptsize brightness}}\tabularnewline[\doublerulesep]
\noalign{\vskip\doublerulesep}
\noalign{\vskip\doublerulesep}
{\scriptsize Planet } & {\scriptsize Scaling factor} & {\scriptsize Detection of $\mathrm{H_{2}O}$ atmosphere}\linebreak {\scriptsize at
$B\approx2000\,\left(>4\,\sigma\right)$ } & {\scriptsize Unambiguous distinction between}\linebreak {\scriptsize clear
$\mathrm{H_{2}O}$ and $\mathrm{H_{2}}$ atmospheres} & {\scriptsize K} & {\scriptsize V}\tabularnewline[\doublerulesep]
\noalign{\vskip\doublerulesep}
\hline 
\noalign{\vskip\doublerulesep}
{\scriptsize GJ~1214b} & {\scriptsize Reference} & {\scriptsize $\sim\unit[80]{ppm}$} & {\scriptsize $\sim\unit[60]{ppm}$} & {\scriptsize 8.78} & {\scriptsize 14.67}\tabularnewline[\doublerulesep]
\noalign{\vskip\doublerulesep}
\noalign{\vskip\doublerulesep}
{\scriptsize HD~97658b} & {\scriptsize 0.092} & {\scriptsize 7.4~ppm} & {\scriptsize 5.5~ppm} & {\scriptsize 5.73} & {\scriptsize 7.71}\tabularnewline[\doublerulesep]
\noalign{\vskip\doublerulesep}
\noalign{\vskip\doublerulesep}
{\scriptsize 55~Cnc~e} & {\scriptsize 0.101} & {\scriptsize 8.1~ppm} & {\scriptsize 6.1~ppm} & {\scriptsize 4.02} & {\scriptsize 5.95}\tabularnewline[\doublerulesep]
\noalign{\vskip\doublerulesep}
\noalign{\vskip\doublerulesep}
{\scriptsize GJ~436b} & {\scriptsize 0.367} & {\scriptsize 29.4~ppm} & {\scriptsize 22.0~ppm} & {\scriptsize 6.07} & {\scriptsize 10.59}\tabularnewline[\doublerulesep]
\hline 
\noalign{\vskip\doublerulesep}
\end{tabular}
\par\end{centering}

\caption{Scaling factors and estimates of required precision in transit depth
measurements to study the atmospheres of HD~97658b, 55~Cnc~e, and
GJ~436b. Stellar and planet parameters are taken from Carter et al.,
2013 for GJ~1214b, \citet{dragomir_new_2013} for HD~97658b, \citet{demory_detection_2011}
and \citet{gillon_improved_2012} for 55~Cnc~e, and \citet{torres_improved_2008}
and \citet{knutson_spitzer_2011-1} for GJ~436b.\label{tab:ScalingForHD97658b}}
\end{table*}

\subsection{Uncertainty Reduction through Stacking Transit Observations\label{sub:Error-Reduction-through}}

In Section 3, we made the assumption that the uncertainty in the spectral
data points scales inversely with the number of transits observed
and photons collected. Based on this assumption, we calculated how
many transit observations need to be stacked with current observational
techniques to detect water absorption in the atmospheres of super-Earths
and to distinguish between hydrogen-rich sub-Neptunes and water worlds.
The assumption that the uncertainty scales inversely with the square
root of the number of observed transits is true for observational
uncertainties that are dominated by white noise, such as photon-noise
or stellar granulation noise. Instrumental effects or long-period
stellar variability may ultimately set a lower limit on achievable
precision for transit observations. To date, however, it is not clear
whether and after how many transits the transit depth uncertainty
will reach a lower limit. 

On the contrary, there is impressive empirical evidence from space-based
observatories such as \textit{Kepler} and \textit{HST} that the transit
depth uncertainty continues to decrease as the number of stacked transits
increases to a few, tens, and even several hundred transits. A prominent
example is the continuous \textit{Kepler} observations of the hot-Jupiter
TrES-2b from which the visible broadband transit depth was determined
to a precision of 1.7~ppm by combining hundreds of transits \citep{barclay_photometrically_2012}.
Spectroscopic observations of three transits of GJ~1214b using \textit{HST
WFC3} by \citet{berta_flat_2012}, reaching a precision within 10\%
of the photon limit, suggest that a similar trend is possible for
spectroscopic transit observations. More observations are required
to fully understand whether the uncertainty of exoplanet transmission
spectra is ultimately limited by instrumental and/or astrophysical
noise.

\subsection{Atmospheric Characterizations of Potentially Habitable Worlds around
M-dwarfs \label{sub:Atmospheric-Characterization-of}}

One compelling feature of the scaling law, Equation (\ref{eq:sigma_req}),
is that the photometric precision required to perform atmospheric
characterizations scales only linearly with the atmospheric temperature
and planet size. The weak scaling illustrates the great potential
of transmission spectroscopy to characterize temperate and small exoplanets
\citep{deming_discovery_2009,seager_exoplanet_2010}. In principle,
transit observations with currently available instrumentation have
the potential to characterize the atmospheres of potentially habitable
super-Earths orbiting nearby M-dwarfs if observational errors continue
to decrease as the number of transits is increased to tens of transits
(see Section \ref{sub:Error-Reduction-through}). 

For example, for a super-Earth planet that orbits a GJ~1214-like
star in the habitable zone at a moderate temperature of $T_{eq}=300\, K$,
the requirements on observational time will rise by only a factor
of $\left(T_{eq,P}/T_{eq,\mathrm{\mathrm{GJ}1214b}}\right)^{2}\approx3.3$
compared to the ones presented in Section 3. Given a sufficiently
large observational program that covers $30-50$ transits, we would,
in principle, have the capability to find water or methane features
with currently available instrumentation, assuming that observational
uncertainty remains to be dominated by the photon noise. Considering
that the orbital period of a temperate planet orbiting a GJ~1214-like
star is only $\sim$9~days, up to 40 transit observations could theoretically
be performed over the course of a year.

\subsection{Distinction Based on Rayleigh Scattering Slope \label{sub:Distinction-based-on}}

Observations of the slope of the Rayleigh scattering at short wavelengths
provide an alternative way of constraining the mean molecular mass.
In practice, however, the Rayleigh slope provides only an upper limit
on the mean molecular mass, thus only a lower limit on the hydrogen
fraction in the atmosphere. The reason is the presence of clouds or
large particle hazes can reduce the observed slope at visible and
UV wavelengths. 

A hydrogen-dominated atmosphere could be identified if a steep negative
slope is observed in the UV-visible part of the transmission spectrum.
If there is no steep slope, however, the distinction between hydrogen-dominated
and volatile-dominated is not possible because the lack of a steep
slope at short wavelength may be due to a high mean molecular mass
or the presence of large particle clouds. The upper limit on the mean
molecular mass can therefore be determined as

\begin{equation}
\mu_{\mathrm{mix}}\lesssim\frac{4k_{B}T}{gR_{*}}\frac{\ln\left(\frac{\lambda_{1}}{\lambda_{2}}\right)}{\left(\frac{R_{p}}{R_{*}}\right)_{\lambda_{2}}-\left(\frac{R_{p}}{R_{*}}\right)_{\lambda_{1}}},\label{eq:rayligh}
\end{equation}

where we derived equation (\ref{eq:rayligh}) from Equation (\ref{eq:1})
using that the Rayleigh cross section $\sigma$ is proportional to
$\lambda^{-4}$ . We assumed that, at least, two measurements of the
transit depth $R_{p}/R_{*}$ are available at wavelengths $\lambda_{1}$
and $\lambda_{2}$ at which Rayleigh scattering dominates, and we
incorporated the uncertainty factor $\left(1\pm\delta T/T\right)$
from Equation (\ref{eq:rayligh}) into the approximately smaller sign.

As an example, we consider a haze-free water-rich scenario on GJ~1214b
in comparison to a range of hydrogen-dominated scenarios with high-altitude
ZnS hazes with different size distributions (Figure \ref{fig:ConstraintsHazes-1}(a)).
In the absence of hazes, the slope of the Rayleigh scattering signature
at short wavelengths can determined from equation (\ref{eq:rayligh})
using an ``approximately equal'' sign instead of the ``approximately
smaller'' sign. In the presence of the ZnS hazes, however, the slope
at short wavelengths depends not only on the mean molecular mass,
but also on the size distribution of the haze particles, the vertical
extent of the particles, and the amount of condensed mass. The slope
is therefore not an unambiguous measure of the mean molecular mass.
For haze particles that are small compared to the wavelength, the
slope of the Rayleigh signature remains unchanged. As the haze particles
become bigger, the slope decreases if the particles are present at
high altitude in sufficient abundance. In the limit of gray, large
particle clouds, more and more of the spectrum becomes flat.

An upper limit on the mean molecular mass can be determined based
on the detection of a straight Rayleigh signature in the UV-visible
spectrum because, at least in the limits of Mie scattering theory
of spherical particles, the opacities of particles do not change with
a slope greater than than $\sigma\propto\lambda^{-4}$ for any realistic
particle size distributions. Besides, we are not aware of any condensate
substances for which changes in the real or imaginary refractive index
at UV-visible wavelengths would steepen the slope across a significant
portion of the UV-visible spectrum. As a result, haze particles would
flatten not steepen the Rayleigh slope at UV-visible transmission
spectrum and we can provide a lower limit, but not an upper limit
on the mean molecular mass.

\section{Summary and Conclusions\label{sec:Summary-and-Conclusions}}

We demonstrated that one can unambiguously distinguish between cloudy
Neptunes with hydrogen atmospheres and low-density super-Earths with
water/volatile-dominated atmospheres by observing the relative sizes
and wing steepnesses of the absorption features in the planet's NIR
transmission spectrum. The proposed observational distinction offers
a promising approach to break the compositional degeneracy in the
interior modeling of low-density super-Earths and Neptunes. We argue
that the distinction can be achieved efficiently for super-Earths
orbiting M-dwarfs by observing water features at moderate spectral
resolution ($R\sim100$) near the brightness peak of the M-dwarf's
spectrum.

In this work, we use the super-Earth GJ~1214b as a case study and
provide a scaling law that scales our quantitative results to other
transiting super-Earths or Neptunes such as HD~97658b, 55~Cnc~e,
and GJ~436b. For GJ~1214b, we show quantitatively that an unambiguous
distinction between cloud-free water-dominated atmospheres and cloudy
hydrogen-dominated atmospheres is possible if the observational uncertainties
can be reduced by a factor of $\sim3$ compared to the published \textit{HST~WFC3}
and \textit{VLT} transit observations by \citet{berta_flat_2012}
and \citet{bean_ground-based_2010}. Similar results can be achieved
using \textit{HST~WFC3 }alone if the observational uncertainty can
be reduced to 35~ppm at $R=70$. 

All results and spectra presented for GJ~1214b are derived using
the new mass, radius, and surface gravity estimates for GJ~1214b
by Carter et al. (2013, in preparation). The 53\% increase in surface
gravity leads to a 35\% decrease in scale height, thereby significantly
altering the strengths of spectral signatures compared to previously published
models.

The required decrease in observational uncertainty may be achievable
for GJ~1214b by observing $10-20$ transits in large programs with
\textit{Hubble} and/or ground-based telescopes. For example, observing
15 transits with \textit{HST WFC3} and increasing the integration
efficiency from 10\% to $\sim$50\% using the new spatial scan mode
on \textit{WFC3} \citep{deming_infrared_2013} would increase the
total number of detected photons by a factor of 25 compared to \citet{berta_flat_2012}.
Assuming that the observations remain dominated by white-noise (see
Section \ref{sub:Error-Reduction-through} for a discussion), such
an observational program would be able to unambiguously distinguish
between a cloud-free water-dominated atmosphere and a cloudy hydrogen-dominated
atmosphere.

The results in this work were obtained using an advanced Bayesian
retrieval framework. The framework not only constrains the atmospheric
parameters, but also determines in a statistically robust way which
molecular species and types of clouds can be inferred from the observational
data.  One main advantage over model-independent approaches to infer
molecular absorbers is that the Bayesian approach inherently accounts
for overlapping spectral features. It assigns a high probability for
the presence of a particular gas or cloud type only if that gas or
cloud type represents the only explanation for the observed data.
The Bayesian approach also enables us to quantify our confidence in
the presence of molecular species if the spectral features are not
box-shaped and are not surrounded by a flat continuum. Spectra of
thick atmospheres generally lack a flat continuum and the signatures
of water vapor and methane resemble characteristic changes in the
transit depth across the whole spectrum rather than distinct features.
Finally, the observational data in the Bayesian framework are analyzed
at full resolution, thereby avoiding any information loss and ambiguity
introduced by binning the data.

The super-Earth GJ~1214b is currently the super-Earth most amenable
to spectroscopic characterizations. The diminutive stellar radius
(0.189~$R_{\odot}$) and the high planetary temperature (>500~K)
result in relatively large transit depth variations if an atmospheric
envelope is present. The atmospheric characterization of other currently
known transiting super-Earths, such as 55~Cnc~e and HD~97658b,
is significantly more challenging because the larger host star radius
decreases the transit signal by a factor of ten or more. 

Large efforts for the discovery of GJ~1214b analogues are, however,
ongoing. Ground-based transit surveys such as \textit{MEarth} \citep{nutzman_design_2008}
currently present the most promising pathway to detect super-Earths
around nearby M-dwarfs that are most amenable to study. The \textit{TESS}
mission expected to launch in 2017 will survey the full sky for planets
transiting nearby M-dwarfs. If an exoplanet like GJ~1214b is found
in the habitable zone of a close-by M-dwarf, this work indicates that
we may have the capability to spectroscopically probe the atmosphere
of a potentially habitable planet - not only in the next decades with
\textit{JWST} or \textit{TPF}-like missions, but today with \textit{HST}
and ground-based telescopes.

We thank Jacob Bean, Zachory Berta, Jean-Michel Désert, and Andras
Zsom for valuable discussions. We thank the anonymous referee for
thorough reading and thoughtful comments that improved the manuscript.

\bibliographystyle{apj}
\addcontentsline{toc}{section}{\refname}\bibliography{MyLibrary}

\begin{thebibliography}{49}
\expandafter\ifx\csname natexlab\endcsname\relax\def\natexlab#1{#1}\fi

\bibitem[{Barclay {et~al.}(2012)Barclay, Huber, Rowe, Fortney, Morley,
  Quintana, Fabrycky, Barentsen, Bloemen, Christiansen, Demory, Fulton,
  Jenkins, Mullally, Ragozzine, Seader, Shporer, Tenenbaum, \&
  Thompson}]{barclay_photometrically_2012}
Barclay, T., {et~al.} 2012, The Astrophysical Journal, 761, 53

\bibitem[{Bean {et~al.}(2010)Bean, Kempton, \&
  Homeier}]{bean_ground-based_2010}
Bean, J.~L., Kempton, E. M.-R., \& Homeier, D. 2010, Nature, 468, 669

\bibitem[{Bean {et~al.}(2011)Bean, D{\'e}sert, Kabath, Stalder, Seager,
  Miller-Ricci~Kempton, Berta, Homeier, Walsh, \& Seifahrt}]{bean_optical_2011}
Bean, J.~L., {et~al.} 2011, The Astrophysical Journal, 743, 92

\bibitem[{Benneke \& Seager(2012)}]{benneke_atmospheric_2012}
Benneke, B., \& Seager, S. 2012, The Astrophysical Journal, 753, 100

\bibitem[{Berta {et~al.}(2011)Berta, Charbonneau, Bean, Irwin, Burke,
  D{\'e}sert, Nutzman, \& Falco}]{berta_gj1214_2011}
Berta, Z.~K., Charbonneau, D., Bean, J., Irwin, J., Burke, C.~J., D{\'e}sert,
  J.-M., Nutzman, P., \& Falco, E.~E. 2011, The Astrophysical Journal, 736, 12

\bibitem[{Berta {et~al.}(2012)Berta, Charbonneau, D{\'e}sert,
  Miller-Ricci~Kempton, {McCullough}, Burke, Fortney, Irwin, Nutzman, \&
  Homeier}]{berta_flat_2012}
Berta, Z.~K., {et~al.} 2012, The Astrophysical Journal, 747, 35

\bibitem[{Brown(2001)}]{brown_transmission_2001}
Brown, T.~M. 2001, The Astrophysical Journal, 553, 1006

\bibitem[{Carter {et~al.}(2011)Carter, Winn, Holman, Fabrycky, Berta, Burke, \&
  Nutzman}]{carter_transit_2011}
Carter, J.~A., Winn, J.~N., Holman, M.~J., Fabrycky, D., Berta, Z.~K., Burke,
  C.~J., \& Nutzman, P. 2011, The Astrophysical Journal, 730, 82

\bibitem[{Charbonneau {et~al.}(2009)Charbonneau, Berta, Irwin, Burke, Nutzman,
  Buchhave, Lovis, Bonfils, Latham, Udry, Murray-Clay, Holman, Falco, Winn,
  Queloz, Pepe, Mayor, Delfosse, \& Forveille}]{charbonneau_super-earth_2009}
Charbonneau, D., {et~al.} 2009, Nature, 462, 891

\bibitem[{Croll {et~al.}(2011)Croll, Albert, Jayawardhana, Kempton, Fortney,
  Murray, \& Neilson}]{croll_broadband_2011}
Croll, B., Albert, L., Jayawardhana, R., Kempton, E. M.-R., Fortney, J.~J.,
  Murray, N., \& Neilson, H. 2011, The Astrophysical Journal, 736, 78

\bibitem[{Crossfield {et~al.}(2011)Crossfield, Barman, \&
  Hansen}]{crossfield_high-resolution_2011}
Crossfield, I. J.~M., Barman, T., \& Hansen, B. M.~S. 2011, The Astrophysical
  Journal, 736, 132

\bibitem[{de~Mooij {et~al.}(2012)de~Mooij, Brogi, de~Kok, Koppenhoefer, Nefs,
  Snellen, Greiner, Hanse, Heinsbroek, Lee, \& van~der
  Werf}]{de_mooij_optical_2012}
de~Mooij, E. J.~W., {et~al.} 2012, Astronomy \& Astrophysics, 538, A46

\bibitem[{Deming {et~al.}(2009)Deming, Seager, Winn, Miller-Ricci, Clampin,
  Lindler, Greene, Charbonneau, Laughlin, Ricker, Latham, \&
  Ennico}]{deming_discovery_2009}
Deming, D., {et~al.} 2009, Publications of the Astronomical Society of the
  Pacific, 121, 952

\bibitem[{Deming {et~al.}(2013)Deming, Wilkins, {McCullough}, Burrows, Fortney,
  Agol, Dobbs-Dixon, Madhusudhan, Crouzet, Desert, Gilliland, Haynes, Knutson,
  Line, Magic, Mandell, Ranjan, Charbonneau, Clampin, Seager, \&
  Showman}]{deming_infrared_2013}
---. 2013, {arXiv:1302.1141}

\bibitem[{Demory {et~al.}(2011)Demory, Gillon, Deming, Valencia, Seager,
  Benneke, Lovis, Cubillos, Harrington, Stevenson, Mayor, Pepe, Queloz,
  S{\'e}gransan, \& Udry}]{demory_detection_2011}
Demory, B.-O., {et~al.} 2011, Astronomy \& Astrophysics, 533, A114

\bibitem[{D{\'e}sert {et~al.}(2011)D{\'e}sert, Kempton, Berta, Charbonneau,
  Irwin, Fortney, Burke, \& Nutzman}]{desert_observational_2011}
D{\'e}sert, J.-M., Kempton, E. M.-R., Berta, Z.~K., Charbonneau, D., Irwin, J.,
  Fortney, J., Burke, C.~J., \& Nutzman, P. 2011, The Astrophysical Journal,
  731, L40

\bibitem[{Dragomir {et~al.}(2013)Dragomir, Matthews, Winn, Rowe, \&
  Team}]{dragomir_new_2013}
Dragomir, D., Matthews, J.~M., Winn, J.~N., Rowe, J.~F., \& Team, M.~S. 2013,
  {arXiv:1302.3321}

\bibitem[{Feroz \& Hobson(2008)}]{feroz_multimodal_2008}
Feroz, F., \& Hobson, M.~P. 2008, Monthly Notices of the Royal Astronomical
  Society, 384, 449

\bibitem[{Feroz {et~al.}(2009)Feroz, Hobson, \&
  Bridges}]{feroz_multinest:_2009}
Feroz, F., Hobson, M.~P., \& Bridges, M. 2009, Monthly Notices of the Royal
  Astronomical Society, 398, 1601{\textendash}1614

\bibitem[{Fortney {et~al.}(2013)Fortney, Mordasini, Nettelmann, Kempton,
  Greene, \& Zahnle}]{fortney_framework_2013}
Fortney, J.~J., Mordasini, C., Nettelmann, N., Kempton, E. M.-R., Greene,
  T.~P., \& Zahnle, K. 2013, A Framework for Characterizing the Atmospheres of
  Low-Mass Low-Density Transiting Planets, {arXiv} e-print 1306.4329

\bibitem[{Gillon {et~al.}(2012)Gillon, Demory, Benneke, Valencia, Deming,
  Seager, Lovis, Mayor, Pepe, Queloz, S{\'e}gransan, \&
  Udry}]{gillon_improved_2012}
Gillon, M., {et~al.} 2012, Astronomy \& Astrophysics, 539, A28

\bibitem[{Gregory(2007)}]{gregory_bayesian_2007}
Gregory, P.~C. 2007, Monthly Notices of the Royal Astronomical Society, 374,
  1321

\bibitem[{Hansen(1971)}]{hansen_multiple_1971}
Hansen, J.~E. 1971, Journal of the Atmospheric Sciences, 28, 1400

\bibitem[{Hansen \& Travis(1974)}]{hansen_light_1974}
Hansen, J.~E., \& Travis, L.~D. 1974, Space Science Reviews, 16, 527

\bibitem[{Heng \& Kopparla(2012)}]{heng_stability_2012}
Heng, K., \& Kopparla, P. 2012, The Astrophysical Journal, 754, 60

\bibitem[{Henry {et~al.}(2011)Henry, Howard, Marcy, Fischer, \&
  Johnson}]{henry_detection_2011}
Henry, G.~W., Howard, A.~W., Marcy, G.~W., Fischer, D.~A., \& Johnson, J.~A.
  2011, {arXiv:1109.2549}

\bibitem[{Hobson \& Jaffe(2009)}]{hobson_bayesian_2009}
Hobson, M.~P., \& Jaffe, A.~H. 2009, Bayesian Methods in Cosmology (Cambridge
  University Press)

\bibitem[{Howe \& Burrows(2012)}]{howe_theoretical_2012}
Howe, A.~R., \& Burrows, A.~S. 2012, The Astrophysical Journal, 756, 176

\bibitem[{Knutson {et~al.}(2011)Knutson, Madhusudhan, Cowan, Christiansen,
  Agol, Deming, D{\'e}sert, Charbonneau, Henry, Homeier, Langton, Laughlin, \&
  Seager}]{knutson_spitzer_2011-1}
Knutson, H.~A., {et~al.} 2011, The Astrophysical Journal, 735, 27

\bibitem[{Kuchner(2003)}]{kuchner_volatile-rich_2003}
Kuchner, M.~J. 2003, The Astrophysical Journal, 596, L105

\bibitem[{Kurokawa \& Kaltenegger(2013)}]{kurokawa_atmospheric_2013}
Kurokawa, H., \& Kaltenegger, L. 2013, Atmospheric mass loss and evolution of
  short-period exoplanets: the examples of {CoRoT-7b} and Kepler-10b, {arXiv}
  e-print 1306.0973

\bibitem[{L{\'e}ger {et~al.}(2004)L{\'e}ger, Selsis, Sotin, Guillot, Despois,
  Mawet, Ollivier, Lab{\`e}que, Valette, Brachet, Chazelas, \&
  Lammer}]{leger_new_2004}
L{\'e}ger, A., {et~al.} 2004, Icarus, 169, 499

\bibitem[{Miller-Ricci {et~al.}(2009)Miller-Ricci, Seager, \&
  Sasselov}]{miller-ricci_atmospheric_2009}
Miller-Ricci, E., Seager, S., \& Sasselov, D. 2009, The Astrophysical Journal,
  690, 1056

\bibitem[{Morley {et~al.}(2013)Morley, Fortney, Kempton, Marley, Visscher, \&
  Zahnle}]{morley_quantitatively_2013}
Morley, C.~V., Fortney, J.~J., Kempton, E. M.-R., Marley, M.~S., Visscher, C.,
  \& Zahnle, K. 2013, Quantitatively Assessing the Role of Clouds in the
  Transmission Spectrum of {GJ} 1214b, {arXiv} e-print 1305.4124

\bibitem[{Morley {et~al.}(2012)Morley, Fortney, Marley, Visscher, Saumon, \&
  Leggett}]{morley_neglected_2012}
Morley, C.~V., Fortney, J.~J., Marley, M.~S., Visscher, C., Saumon, D., \&
  Leggett, S.~K. 2012, The Astrophysical Journal, 756, 172

\bibitem[{Nettelmann {et~al.}(2011)Nettelmann, Fortney, Kramm, \&
  Redmer}]{nettelmann_thermal_2011}
Nettelmann, N., Fortney, J.~J., Kramm, U., \& Redmer, R. 2011, The
  Astrophysical Journal, 733, 2

\bibitem[{Nutzman \& Charbonneau(2008)}]{nutzman_design_2008}
Nutzman, P., \& Charbonneau, D. 2008, Publications of the Astronomical Society
  of the Pacific, 120, 317, {ArticleType:} research-article / Full publication
  date: March 2008 / Copyright {\textcopyright} 2008 The University of Chicago
  Press

\bibitem[{Querry(1987)}]{querry_optical_1987}
Querry, M. 1987, 333

\bibitem[{Rogers \& Seager(2010{\natexlab{a}})}]{rogers_framework_2010}
Rogers, L.~A., \& Seager, S. 2010{\natexlab{a}}, The Astrophysical Journal,
  712, 974

\bibitem[{Rogers \& Seager(2010{\natexlab{b}})}]{rogers_three_2010}
---. 2010{\natexlab{b}}, The Astrophysical Journal, 716, 1208

\bibitem[{Seager \& Deming(2010)}]{seager_exoplanet_2010}
Seager, S., \& Deming, D. 2010, Annual Review of Astronomy and Astrophysics,
  48, 631

\bibitem[{Seager {et~al.}(2007)Seager, Kuchner, {Hier-Majumder}, \&
  Militzer}]{seager_massradius_2007}
Seager, S., Kuchner, M., {Hier-Majumder}, C.~A., \& Militzer, B. 2007, The
  Astrophysical Journal, 669, 1279

\bibitem[{Sellke {et~al.}(2001)Sellke, Bayarri, \&
  Berger}]{sellke_calibration_2001}
Sellke, T., Bayarri, M.~J., \& Berger, J.~O. 2001, The American Statistician,
  55, 62

\bibitem[{Sivia \& Skilling(2006)}]{sivia_data_2006}
Sivia, D.~S., \& Skilling, J. 2006, Data analysis: a Bayesian tutorial (Oxford
  University Press)

\bibitem[{Skilling(2004)}]{skilling_nested_2004}
Skilling, J. 2004, {AIP} Conference Proceedings, 735, 395

\bibitem[{Teske {et~al.}(2013)Teske, Turner, Mueller, \&
  Griffith}]{teske_optical_2013}
Teske, J.~K., Turner, J.~D., Mueller, M., \& Griffith, C.~A. 2013, Monthly
  Notices of the Royal Astronomical Society, 431, 1669

\bibitem[{Torres {et~al.}(2008)Torres, Winn, \& Holman}]{torres_improved_2008}
Torres, G., Winn, J.~N., \& Holman, M.~J. 2008, The Astrophysical Journal, 677,
  1324

\bibitem[{Trotta(2008)}]{trotta_bayes_2008}
Trotta, R. 2008, Contemporary Physics, 49, 71

\bibitem[{Winn(2011)}]{winn_transits_2011}
Winn, J.~N. 2011, in Exoplanets (University of Arizona Press)

\end{thebibliography}

\end{document}